\documentclass[reprint,aps,prd,notitlepage,superscriptaddress,nofootinbib]{revtex4-2}

\usepackage[utf8]{inputenc}
\usepackage[english]{babel}
\usepackage{amsmath}
\usepackage{amsfonts}
\usepackage{amssymb}
\usepackage{listings}
\usepackage{lipsum}
\usepackage{multirow}
\usepackage{datetime}
\usepackage{graphicx}
\usepackage{mathtools}
\usepackage{mathrsfs}
\usepackage{dcolumn}
\usepackage{multirow}
\usepackage{subfig}
\usepackage{soul}
\usepackage{breqn}
\usepackage{chngcntr}
\usepackage{color} 
\usepackage{hyperref}
\hypersetup{
    colorlinks=true, 
    pdfborder = {0 0 0.5 [3 3]},
    anchorcolor=black,
    citecolor=blue,
    linktoc=all,    
    linktocpage=true,
    linkcolor=red,
	urlcolor=blue
}

\begin{document}

\title{Self-gravity effects of ultralight boson clouds formed by black hole superradiance}

\author{Taillte May}
\email{tmay@perimeterinstitute.ca}
\affiliation{Perimeter Institute for Theoretical Physics\char`,{} Waterloo\char`,{} Ontario N2L 2Y5\char`,{} Canada}
\affiliation{Department of Physics \& Astronomy\char`,{} University of Waterloo\char`,{} Waterloo\char`,{} Ontario N2L 3G1 Canada}

\author{William E. East}
\email{weast@perimeterinstitute.ca}
\affiliation{Perimeter Institute for Theoretical Physics\char`,{} Waterloo\char`,{} Ontario N2L 2Y5\char`,{} Canada}

\author{Nils Siemonsen}
\email{nils.siemonsen@princeton.edu}
\affiliation{Princeton Gravity Initiative\char`,{} Princeton University\char`,{} Princeton New Jersey 08544\char`,{} USA}
\affiliation{Department of Physics\char`,{} Princeton University\char`,{} Princeton\char`,{} New Jersey 08544\char`,{} USA}

\date{\today}

\begin{abstract} 
Oscillating clouds of ultralight bosons can grow around
spinning black holes through superradiance, extracting energy and angular
momentum, and eventually dissipating through
gravitational radiation. Gravitational wave detectors like LIGO, Virgo, KAGRA, and LISA
can thus probe the existence of ultralight bosons.
In this study, we use fully general-relativistic
solutions of the black hole-boson cloud systems to study the self-gravity effects
of scalar and vector boson clouds, making only the simplifying assumption that
the spacetime is axisymmetric (essentially corresponding to taking an oscillation average).
We calculate the self-gravity shift in the cloud oscillation frequency, which
determines the frequency evolution of the gravitational wave signal,
finding that this effect can be up to twice as large in the relativistic
regime compared to nonrelativistic estimates. 
We use this to improve the \texttt{superrad} waveform model,
and estimate that this reduces the theoretical phase error to a few cycles over the characteristic timescale of the gravitational
wave emission timescale for the louder vector boson signals. 
We also perform an analysis of the spacetime geometry of these systems,
calculating how the cloud changes the innermost stable circular orbit and light ring around the black
hole.
\end{abstract}

\maketitle

\section{Introduction}
An intriguing possibility of recent focus is the prospect of using
observations of black holes and gravitational waves to probe the existence of
ultralight bosons.  There are several theoretical motivations for considering
scalar and vector particles that are weakly coupled to the Standard Model. 
The quantum chromodynamics (QCD) axion is a (pseudo) scalar example
\cite{Peccei:1977hh,Weinberg:1977ma}, whose mass is observationally constrained
to $m_b \leq 10^{-3}$ eV \cite{Caputo:2024oqc}, and for some models is
theoretically expected to be $m_b \sim 10^{-10}$ eV
\cite{Kim:2008hd,Arvanitaki:2010sy,Hui:2016ltb}. Axions, along with dark
photons~\cite{Holdom:1985ag,Pani:2012bp,Graham:2015rva,Agrawal:2018vin}, have
been considered as potential dark matter candidates; string theory also
features ultralight axionlike particles \cite{Arvanitaki:2009fg}, as well as
vector bosons \cite{Goodsell:2009xc} in the mass range $10^{-33}$ to $10^{-10}$ eV. 

Black holes can produce ultralight bosons through the
superradiant mechanism.  A scattering field can extract rotational energy
from a black hole if it satisfies the superradiance condition,
$\omega<m\Omega_H$, where $\omega$ is the angular frequency of the field, $m$ is
its azimuthal mode number, and $\Omega_H$ is frequency of the black hole
horizon~\cite{1971JETPL..14..180Z,Misner:1972kx,Starobinsky:1973aij}.
A field in a bound state that satisfies the superradiance condition will be
continuously amplified through this process, and grows exponentially with time.
In the absence of other couplings, the resulting boson cloud eventually spins
down the black hole sufficiently to saturate the superradiance condition; after
this, there is minimal flux through the black hole horizon and the cloud will 
dissipate through gravitational radiation.
This process is possible in principle for a bosonic field with any mass, though
the instability rate is only astrophysically relevant if the Compton wavelength of the
field is roughly comparable to the black hole's radius. This is equivalent to $\alpha
\equiv G m_b M_{\text{BH}}/(c^3 \hbar)\sim 1$, where $m_b$ is the boson's mass
and $M_{\rm BH}$ the black hole's mass.

The black hole superradiant instability occurs for 
minimally coupled ultralight scalar and vector bosons%
\footnote{
The superradiant instability can also occur
for massive spin-2 fields~\cite{Brito:2013wya,Dias:2023ynv}, though the observational implications
are complicated by the fact that in this case there is 
another instability that occurs even for nonspinning black holes~\cite{Babichev:2013una}.
This instability is even faster in much 
of the parameter space and the (potentially model-dependent) nonlinear evolution
has only begun to be explored~\cite{East:2023nsk}. 
}, therefore, it can probe particles with arbitrarily weak couplings
to Standard Model particles, complementing terrestrial experiments.
Black hole spin measurements and gravitational waves from binary black hole mergers can be used to place constraints on the existence of ultralight bosons \cite{Arvanitaki:2014wva,Arvanitaki:2016qwi,Cardoso:2018tly,Ng:2020ruv,Baryakhtar:2017ngi,Stott:2020gjj}.
The gravitational waves emitted by a dissipating
boson cloud fall in the LIGO-Virgo-KAGRA sensitivity band \cite{LIGOScientific:2014pky,VIRGO:2014yos,Aso:2013eba,KAGRA:2020tym}
for boson masses in the range from $10^{-13}$
to $10^{-11}$ eV, while boson masses in between $10^{-18}$ and 
$10^{-15}$ eV correspond to LISA's future sensitivity range \cite{LISA:2017pwj}. 
Stochastic background searches and blind continuous wave searches can be used
to look for signals for some unknown population of black holes that has
undergone the superradiant instability
\cite{LIGOScientific:2021rnv,Tsukada:2018mbp,Tsukada:2020lgt,Palomba:2019vxe,KAGRA:2022osp,Brito:2017zvb,Arvanitaki:2010sy,Arvanitaki:2014wva,Yoshino:2014wwa,Arvanitaki:2016qwi,Brito:2017wnc,Cardoso:2018tly,Zhu:2020tht},
though placing constraints in the absence of a detection requires one to make
assumptions about the (largely unknown) black hole population. Alternatively, one can consider
targeting specific known black holes \cite{Sun:2019mqb}. In particular, performing follow-up
searches of the merger remnants of observed binary black hole mergers has the
advantage that the properties of a merger remnant can be estimated from the
merger signal, and so specific boson masses can be constrained with minimal 
assumptions~\cite{Isi:2018pzk,Chan:2022dkt,Jones:2023fzz}. However, this is feasible
only for the fastest evolving boson clouds and loudest gravitational wave signals.
For these, it is important to have a relativistically accurate gravitational waveform model,
a crucial aspect of which we develop in this work.

The gravitational wave signal from a superradiant cloud is mostly monochromatic,
but with a small upward drift in the frequency due to the changing mass of the
boson cloud. Not accounting for this limits the timescales over which the
signal can be coherently tracked by gravitational wave search methods. 
The frequency shift due to the self-gravity of the cloud was first calculated in
Refs.~\cite{Baryakhtar:2017ngi} and \cite{Isi:2018pzk,Baryakhtar:2020gao}, in the 
Newtonian limit of the fastest growing vector and scalar modes, respectively.  
In Ref.~\cite{Siemonsen:2022yyf}, this was extended to higher azimuthal modes and
compared to a quasirelativistic calculation introduced in Ref.~\cite{Siemonsen:2019ebd}.
The latter is still based on evaluating the
Newtonian expressions for the frequency shift, but employs relativistic 
test-field solutions to calculate the energy density. More recently, Ref.
\cite{Cannizzaro:2023jle} used relativistic perturbation theory to estimate
this effect, but still treating the perturbation from the cloud's 
self-gravity in terms of a Newtonian potential, and found an improvement from the nonrelativistic
limit. 

As we show below, these previous nonrelativistic estimates for the frequency shift
can be off by an order-one factor in the relativistic regime.  In this work, we account for the
self-gravity effects of the boson cloud around a black hole in a fully
relativistic way by leveraging stationary spacetime solutions describing
black holes with synchronized complex scalar or vector clouds, first
constructed in
Refs.~\cite{Herdeiro:2014goa,Herdeiro:2015gia,Herdeiro:2016tmi,Santos:2020pmh}.
We utilize these to determine the cloud mass dependence of the cloud frequency, and
hence, the frequency of the gravitational wave signal in an oscillation-average
approximation for $m=1$ scalar and vector boson clouds. Using this, we update
the waveform model \texttt{SuperRad}\footnote{
Available at \href{https://www.bitbucket.org/weast/superrad/}{https://www.bitbucket.org/weast/superrad}.
}, introduced in
Ref.~\cite{Siemonsen:2022yyf}. With these results, the accumulated phase error is
reduced by several orders of magnitude, and for vector bosons is on the 
order of one radian or smaller over the characteristic timescale of the gravitational wave
signal in much of the parameter space.

The black hole-boson cloud solutions can also be used to determine how the cloud changes 
the black hole's spacetime, and hence, how it affects the redshift, lensing, shadows, etc. associated
with the system, as studied in Refs.~\cite{Herdeiro:2015gia,Ni:2016rhz,Sengo:2022jif}. Here,  
we compare the geometry of the scalar cloud case to the vector case in terms
the effect on quantities like the innermost stable circular orbit (ISCO) radius 
and light ring, and comment on how this may affect measurements of black hole spin.

The layout of this paper is as follows. In Sec. \ref{sec:preliminaries}, we
provide some background on the nonrelativistic treatment of this problem, and
some details on the oscillation-averaged approximation considered here. We describe
how we numerically construct solutions in Sec. \ref{sec:methods}, with more details included in
the Appendices. We present our main results in Sec. \ref{sec:results}, showing
the relativistic correction to the frequency in Sec. \ref{sec:frequency_shift},
with an updated estimate of the
phase error in the superradiance waveform in Sec.
\ref{sec:implications_bosons}, 
as well as the effect of the presence of a boson cloud on the geometry in Sec.
\ref{sec:geometry}. Finally, we conclude with a discussion of the implications of
these results in Sec. \ref{sec:discussion}. In the following, we use
geometrized units ($G=c=1$) and the unit-adjusted boson mass $\mu \equiv m_b /
\hbar$.

\section{Preliminaries}
\label{sec:preliminaries}

\subsection{Nonrelativistic superradiance}

In this section, we review the solutions describing boson clouds arising from
superradiance in the nonrelativistic limit and their associated timescales.
This is to provide context for the results of this study, as well as to
introduce some useful definitions. 

First, we note that the nonrelativistic limit for black hole superradiance
corresponds to the regime $\alpha \equiv M_{\text{BH}}\mu \ll 1$, where the
characteristic length scale of the black hole is much smaller than that of the
cloud [the former being given by $\sim\alpha/\mu$ and the latter by $\sim 1/(\alpha \mu$)]. 
In the nonrelativistic limit, the governing test-field equations can be solved
analytically \cite{Detweiler:1980uk,Yoshino:2013ofa,Brito:2014wla,Baryakhtar:2017ngi} to obtain
a spectrum of superradiantly unstable modes. 
These modes are analogous to the energy states in the hydrogen atom, and therefore,
are also labeled by the numbers $\ell$, $m$, and $n$,
corresponding to the orbital angular number, azimuthal number, and number of radial
nodes. 
The frequencies of the bound states for a boson field
around a black hole are given by
\begin{align}
    \omega \approx \mu \left(1-\frac{\alpha^2}{2(n+\ell+1)^2}+\mathcal{O}(\alpha^3)\right).
\end{align}
In this study, we will focus on the $(\ell,m,n) = (1,1,0)$ mode for the scalar case
and the $(\ell,m,n) = (0,1,0)$ for the vector case, which correspond to the
fastest growing modes at the respective spins.
The cloud growth timescale for the fastest growing mode, to leading order in $\alpha$, scales as \cite{Detweiler:1980uk}
\begin{align}
    \tau_{\text{inst.}}^{\text{sca}} \sim \frac{1}{\alpha^{9}(m\Omega_H-\omega)}
    ,
\end{align}
in the scalar case. In the vector case the fastest growing mode has a growth timescale that scales as \cite{Baryakhtar:2017ngi}
\begin{align}
    \tau_{\text{inst.}}^{\text{vec}} \sim \frac{1}{\alpha^{7}(m\Omega_H-\omega)},
\end{align}
where in this limit $\omega \approx \mu \ll m\Omega_H$.
Once the cloud extracts a sufficient amount of angular momentum from the black hole, the superradiance condition is nearly saturated $\Omega_H \approx \omega/m$, and the cloud dissipates through gravitational radiation. The emission timescale $\tau_{\text{GW}}$ is the time for the cloud to decrease to
half its original mass through gravitational dissipation after saturation.
This emission timescale for the fastest growing (and dissipating) modes scale with $\alpha$ to leading order as follows \cite{Brito:2014wla,Baryakhtar:2017ngi}:
\begin{align}
    %&\tau_{\text{GW}}^{\text{sca}} \propto \frac{1}{\alpha^{14}},
    %&\tau_{\text{GW}}^{\text{vec}} \propto \frac{1}{\alpha^{10}}
    &\tau_{\text{GW}}^{\text{sca}} \sim \alpha^{-14}\left ( \frac{M_{\rm BH}^2}{M_{\rm cloud}}\right)_{t=t_{\rm sat}} ,\nonumber\\
    &\tau_{\text{GW}}^{\text{vec}} \sim \alpha^{-10}\left ( \frac{M_{\rm BH}^2}{M_{\rm cloud}} \right)_{t=t_{\rm sat}} ,
\end{align}
where the mass of the cloud $M_{\rm cloud}$ is evaluated at the time it reaches its maximum. 
Note that $\tau_{\text{inst}}\ll\tau_{\text{GW}}$.
As indicated by the above scalings, vector boson clouds can grow and dissipate on much shorter timescales than scalar clouds. 

\subsection{Nonrelativistic frequency shift}
\label{sec:non-rel_shift_calc}

Generally, there will be a contribution to the gravitational binding energy due to the boson cloud's self-
gravity, which was neglected in the above expressions. Most notably, the cloud's frequency $\omega$ (and hence, also the gravitational wave frequency) is shifted by an amount dependent on the cloud's mass $M_{\rm cloud}$.
We define $\Delta \omega = \omega(M_{\text{cloud}}) - \omega(M_{\text{cloud}}=0)$ to be this cloud mass dependent frequency shift. 
As the cloud dissipates through gravitational wave emission and $M_{\rm{cloud}}$ decreases, this leads to a 
gradual increase of the gravitational wave frequency over time.
Following Refs.~\cite{Baryakhtar:2017ngi,Isi:2018pzk,Baryakhtar:2020gao,Siemonsen:2022yyf}, this frequency shift was determined in the nonrelativistic regime. There, cloud's energy density $\rho(\textbf{r})$ sources the Newtonian potential
\begin{align}
    U(\mathbf{r}) = -\int d^3\mathbf{r'}\frac{\rho(\mathbf{r'})}{\left|\mathbf{r}-\mathbf{r'}\right|}
    .
\end{align}
The leading order in $\alpha$ expression for the energy density of the cloud $\rho(\mathbf{r})$ [where, in this limit, the cloud mass is $M_{\text{cloud}} = \int d^3\mathbf{r} \rho(\mathbf{r})$] is given by $\rho(\mathbf{r}) = \mu^2 \psi(\mathbf{r})^2$ and $\rho(\mathbf{r}) = \mu^2 \mathcal{A}_a(\mathbf{r}) \mathcal{A}^{a}(\mathbf{r})$, using the analytic nonrelativistic solution of the cloud distributions $\psi(\mathbf{r})$ and $\mathcal{A}^a(\mathbf{r})$ from, e.g., Refs.~\cite{Brito:2014wla,Baryakhtar:2017ngi}. Here and in the following, we use $\mu$ interchangeably for the scalar and vector masses (as done also for $\omega$ and $m$). Note also, we dropped the time-dependencies of $\psi(\mathbf{r})$ and $\mathcal{A}^a(\mathbf{r})$, since these integrate away as discussed below. The frequency shift is then 
\begin{align}
    \Delta\omega\frac{M_{\text{cloud}}}{\mu} &\approx \int d^3\mathbf{r} \rho(\mathbf{r})U(\mathbf{r}) \nonumber\\
    &= -2\int d^3 \mathbf{r}\int_{|\mathbf{r'}|<|\mathbf{r}|}d^3\mathbf{r'}\frac{\rho(\mathbf{r})\rho(\mathbf{r'})}{\left|\mathbf{r}-\mathbf{r'}\right|} = 2W
    ,
    \label{eq:bindingE}
\end{align}
where $W$ is the nonrelativistic expression for the binding energy of the cloud%
\footnote{N.B. the factor of two difference in Eq.~\eqref{eq:bindingE} compared to some previous references, 
as discussed in Ref.~\cite{Siemonsen:2022yyf}.}.
We can then perform a multipole expansion for the denominator in Eq. \eqref{eq:bindingE}. When $|\mathbf{r}|\geq |\mathbf{r}'|$, we have that
\begin{align} \label{multipole}
    \frac{1}{|\mathbf{r}-\mathbf{r}'|} = \sum_{l=0}^{\infty} \sum_{m=-l}^{l}\left( \frac{|\textbf{r}'|^{l}}{|\textbf{r}|^{l+1}}\right)\frac{4\pi}{2l+1}Y_{lm}(\Omega)Y^*_{lm}(\Omega')
    ,
\end{align}
where $Y_{l m}(\Omega)$ are the spherical harmonics, $\Omega$ denotes the two angles on the sphere, and ${}^*$ indicates complex conjugation.
In the monopole approximation, where only the first term ($l = 0$) in this series is included, Eq. \eqref{eq:bindingE} gives
\begin{align} \label{Monopole}
    \Delta\omega \approx -\frac{2\mu}{M_{\text{cloud}}}\int d^3\mathbf{r} \int_{|\mathbf{r'}|<|\mathbf{r}|} d^3\mathbf{r'} \frac{\rho(\mathbf{r})\rho(\mathbf{r'})}{|\mathbf{r}|}  
    .
\end{align} 
In Ref.~\cite{Siemonsen:2022yyf}, several higher multipoles were calculated and the error in the monopole 
approximation (when considered the leading order in $\alpha$ term) is found to be at the percent level in the scalar case, and subleading in $\alpha$ in the vector case. As noted above, after performing this integral, the time dependence of the potential drops out.
The final result in the monopole and nonrelativistic approximation is~\cite{Siemonsen:2022yyf}
\begin{align}
    \Delta\omega^{\text{sca}}_{\text{nonrel}} &\approx -\frac{93}{512}\frac{M_{\text{cloud}}\alpha^3}{M_{\text{BH}}^2},\\
    \Delta\omega^{\text{vec}}_{\text{nonrel}} &\approx -\frac{5}{8}\frac{M_{\text{cloud}}\alpha^3}{M_{\text{BH}}^2}.
\end{align}

\subsection{Relativistic frequency shift}
\label{sec:osc_avg}

The aim of this work is to calculate the self-gravity effects of the cloud,
including the frequency shift, in a relativistically accurate way. In principle,
this could be calculated using black hole perturbation theory by solving for the
metric perturbation due to the cloud, and then determining how this shifts the
eigenfrequencies of the boson cloud solutions.  However, given the intricacies
of beyond-leading order black hole perturbation theory, here we opt instead to
numerically solve the full governing Einstein equations. For a single (real) bosonic field, the
presence of the cloud means that the spacetime is no longer exactly stationary
or axisymmetric, so solving for the full, dynamical spacetimes while covering
the parameter space (with the accuracy we require) would be infeasible
without reducing the dimensionality of the problem.
Instead, we calculate the self-gravity effects of the cloud in a related
axisymmetric and stationary spacetime, where instead of considering a real
scalar or vector field, we consider a complex field with a harmonic time and azimuthal dependence
given by $\sim e^{i(m\varphi - \omega t)}$. This is equivalent to considering a superposition
of two real bosonic fields (with identical masses) with a relative phase shift
of $\pi/2$. In this sense, considering a stationary complex field instead of
a real one is an oscillation-averaged approximation.  When we further assume
that the bosonic field saturates the black hole superradiance condition exactly
($\omega = m\Omega_H$), and thus has no flux through the horizon, then one
obtains the solutions found in Refs.~\cite{Herdeiro:2014goa,Herdeiro:2015gia,Herdeiro:2016tmi,Santos:2020pmh}. 

Applying results from these stationary solutions to the gravitational
wave-emitting cases is justified by the fact that the gravitational radiation
timescale is always much longer than the oscillation period, and the energy
density of the boson cloud solutions (in the Newtonian limit) is axisymmetric
to a large degree. However,
this does introduce a small theoretical error.  We explicitly estimate this
in Appendix \ref{sec:complex_err_newtonian_term},
and include this in our overall error estimates, as discussed below.
We also compare the relativistic frequency calculation 
to the frequency obtained in 
a numerical relativity evolution without symmetry assumptions in Appendix~\ref{sec:nr_comp}.
We find the two are consistent, though the uncertainties in latter are significantly
larger than the expected error in the former.

\section{Methods}
\label{sec:methods}

\subsection{Construction of solutions}

As in Refs.~\cite{Herdeiro:2014goa,Herdeiro:2015gia,Herdeiro:2016tmi,Santos:2020pmh}, we numerically construct axisymmetric, stationary spacetimes describing a black hole with a (synchronized) boson cloud. To that end, we solve the fully relativistic Einstein-Klein-Gordon and Einstein-Proca equations for scalar and vector boson clouds, respectively. These are, for a complex massive scalar field $\Phi$ coupled to gravity, 
\begin{align}
    G_{a b}/(8 \pi) &= \nabla_b \Phi^* \nabla_a\Phi+ \nabla_a \Phi^* \nabla_b\Phi \nonumber \\ &-g_{ab}\left(\mu^2\Phi^*\Phi+\nabla_c\Phi\nabla^c\Phi^*\right),\\
    \nabla_a\nabla^a \Phi  &= \mu^2 \Phi
    ,
\end{align}
where $G_{ab}$ is the Einstein tensor associated with the metric $g_{ab}$. For the case of a complex massive vector field $A_b$ with 
associated field strength tensor 
\begin{align}
    F_{ab} &= \nabla_aA_b-\nabla_bA_a,
\end{align}
the Einstein-Proca equations are
\begin{align}
    G_{a b}/(8 \pi) &= \frac{1}{2}\mu^2\left(A_bA^*_a+A_aA^*_b\right)-\frac{1}{2}\mu^2 A_cA^{*c}g_{ab}\nonumber \\&-\frac{1}{4}g_{ab}F_{cd}^*F^{cd}-\frac{1}{2}g^{cd}\left(F^*_{ac}F_{db}+F^*_{bc}F_{da}\right),\\
    \nabla^a F_{ab}&=\mu^2A_b
    .
\end{align}
We do not consider any additional nongravitational interactions.
Imposing axisymmetry and stationarity of the metric, the spacetime
exhibits the Killing vectors $\varphi_a$ and $t_a$, respectively.
This assumption implies that the Lie derivative of the scalar field
along the Killing vector fields satisfies
\begin{align}
    \mathcal{L}_{\varphi} \Phi = i m \Phi, \quad \mathcal{L}_{t} \Phi = -i \omega \Phi , 
\end{align}
and similarly for the vector $A_a$. We further assume the
synchronicity condition between the cloud frequency and the black hole horizon,
$\omega = m\Omega_H $; i.e., the bosonic fields saturate the black hole
superradiance condition. In this study,
we restrict to solutions corresponding to the endpoint of the fastest growing
superradiant unstable modes: we set $m=1$.

Our general approach to numerically constructing these solutions is similar to
Refs.~\cite{Herdeiro:2014goa,Herdeiro:2015gia,Herdeiro:2016tmi,Santos:2020pmh},
though the particular form for the metric ansatz that we use and some other
details are different, and we give the full expressions for the elliptic
equations that are solved in Appendix \ref{sec:eom}.  As in
Refs.~\cite{Herdeiro:2014goa,Herdeiro:2015gia,Herdeiro:2016tmi,Santos:2020pmh},
we use a grid covering the angular domain and a compactified radial coordinate
that goes from the black hole horizon to spatial infinity. 
We solve these discretized partial differential equations using an adapted
version of Newton-Raphson method that was initially developed for boson star
initial data in Ref.~\cite{Siemonsen:2020hcg}. The resulting data is publicly available at \cite{dataref}. For more details on how these
solutions are constructed, see Appendix \ref{sec:eom}.

\subsection{Covering the parameter space}
\label{sec:fitting_data}
We construct black hole-boson cloud solutions across the two dimensional
parameter space labeled by the frequency $\omega$ and the black hole radius 
(in units of the boson mass $\mu$) by beginning from some seed solution and
making small changes to these input quantities to relax to a nearby solution
and then repeating.  For each solution, we record the frequency $\omega$, the
black hole mass $M_{\text{BH}}$ and angular momentum $J_{\text{BH}}$, and the
total boson cloud mass $M_{\text{cloud}}$ and angular momentum
$J_{\text{cloud}}$ (see Appendix~\ref{sec:physical_quantities} for how these quantities are
calculated).  The two
dimensional parameter space of solutions can alternatively be parametrized in
terms of $M_{\text{cloud}}/M_{\text{BH}}$ and $\alpha$. In terms of cloud mass, we cover
the range going from a maximum of $M_{\text{cloud}}/M_{\text{BH}} \approx 0.12$
down to small values of the cloud mass (as low as $M_{\text{cloud}}/M_{\text{BH}} \approx 0.001$, 
which can then be linearly extrapolated to zero). These encompass the
cloud solutions that might arise through superradiance\footnote{We estimate whether a
cloud could be formed around an isolated black hole via superradiance by
checking whether the total angular momentum in the spacetime divided by the
total mass squared is less than one.  This gives that
$M_{\text{cloud}}/M_{\text{BH}}< 0.11$.}. In terms of $\alpha$, we consider
values ranging from the maximum allowed values for $m=1$ ($\alpha\approx 0.5$
for scalars and $\alpha\approx0.6$ for vectors), down to small values of
$\alpha$, where the scales of this problem become disparate [recall, the black hole radius is $\sim
\alpha/ \mu$, while the cloud radius is $\sim 1/(\alpha\mu)$],
requiring more computational resources to resolve both objects. Due to this
limitation, we include results for numerical solutions with $\alpha >
\alpha_{\text{min}}$, where $\alpha_{\text{min}} = 0.2$ in the scalar case and
$\alpha_{\text{min}} = 0.09$ in the vector case. We reach a lower value of
$\alpha$ in the vector case because the fastest growing state is more compact
than in the scalar case. In total, we compute $\approx 6000$ solutions in the scalar case and $\approx 13000$ solutions in the vector case.

We use these numerical solutions to interpolate and extrapolate over the whole
physically relevant parameter space. 
First, a bivariate spline is used to interpolate over these solutions%
\footnote{We use a fifth order bivariate spline interpolation algorithm from \texttt{scipy}.
} and calculate the frequency on a regular grid of $\alpha$ and $M_{\text{cloud}}/M_{\rm BH}$. 
For each fixed value of $\alpha$, we then fit $\omega$ to the best fit polynomial in cloud mass 
of the form
\begin{align}
    \label{eq:full_numeric_fit}
    \frac{\omega}{\mu} = \sum_{q = 0}^{2} f_q(\alpha) \left(\frac{M_{\text{cloud}}}{M_{\text{BH}}}\right)^q
    ,
\end{align}
where the choice to include only up to quadratic terms in cloud mass 
is motivated by the negligible contribution of higher order terms, as detailed in Appendix \ref{sec:higher_order_mc_fit}.
Then the cloud mass-independent part of the frequency is described by $f_0$, and the frequency shift $\Delta\omega = \omega - \omega(M_{\text{cloud}}=0)$ is described by $f_1$ and $f_2$ for each value of $\alpha$. 
We compare $f_0(\alpha)$ with test field results and find good agreement (see Appendix \ref{sec:consistency_checks} for details). Then we interpolate\footnote{Here we use a cubic spline interpolating algorithm from \texttt{scipy}.} $f_q(\alpha)$ over $\alpha$ to return a smooth function that can be evaluated at any $\alpha$.

Extrapolation of these relativistic results to lower values of $\alpha<\alpha_{\text{min}}$ is done by
leveraging the nonrelativistic results reviewed in Sec.~\ref{sec:non-rel_shift_calc}. We fit the difference in the coefficients in Eq.~\eqref{eq:full_numeric_fit} from
their nonrelativistic limit with a higher order polynomials of the form

\begin{align}
f_1(\alpha) &+ C_{\text{nr}}\alpha^2 = \sum_{p=3}^6 c_{p,1} \alpha^p,\nonumber  \\
    \label{eq:extrapolation_shift_fit}
     f_2(\alpha) &= \sum_{p=3}^6 c_{p,2} \alpha^p 
     ,
\end{align}
where $C_{\text{nr}}$ is calculated from the nonrelativistic limit (see Sec. \ref{sec:non-rel_shift_calc}), with $C_{\text{nr}}=93/512$ in the scalar case and $C_{\text{nr}}=5/8$ in the vector case. We fit for $c_{p,q}$ using points in the low $\alpha$ regime\footnote{We use points with $\alpha<0.3$ in the scalar case and $\alpha<0.15$ in the vector case.} and extrapolate to lower values to estimate the relativistic correction there. We find that the fit parameters in Eq.~\eqref{eq:extrapolation_shift_fit} are order unity or smaller, guaranteeing that the fit will approach the expected nonrelativistic limit at small $\alpha$.

\section{Results}
\label{sec:results}

\subsection{Frequency shift}
\label{sec:frequency_shift}

\begin{figure}
    \centering   
    \subfloat[\centering Scalar field]{{\includegraphics[width=8cm]{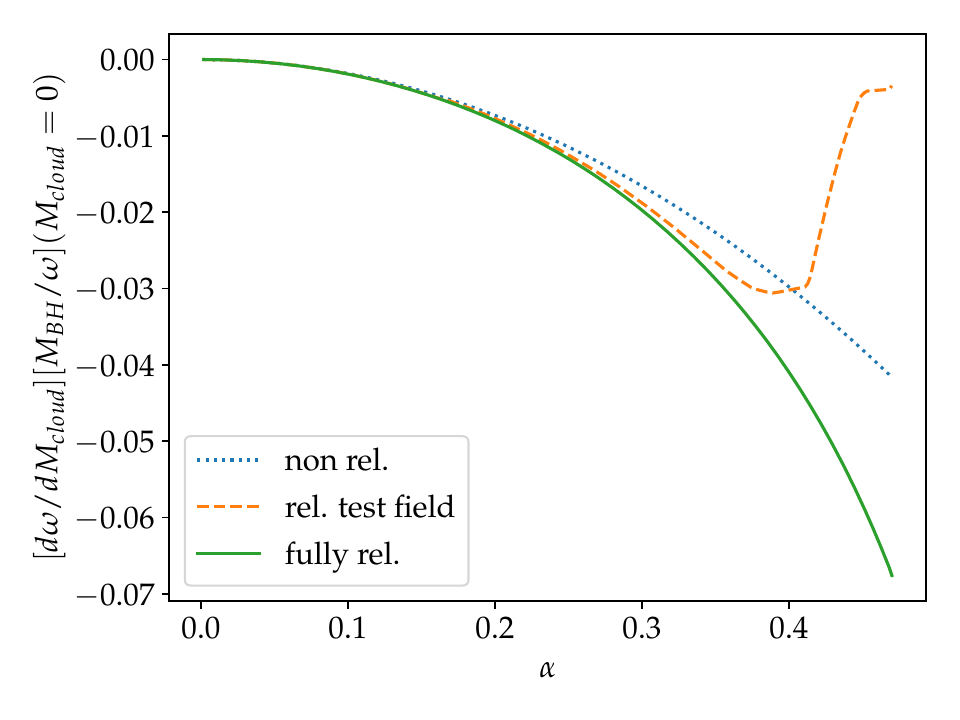} \label{fig:shiftValpha_a}}}%
    \qquad
    \subfloat[\centering Vector field]{{\includegraphics[width=8cm]{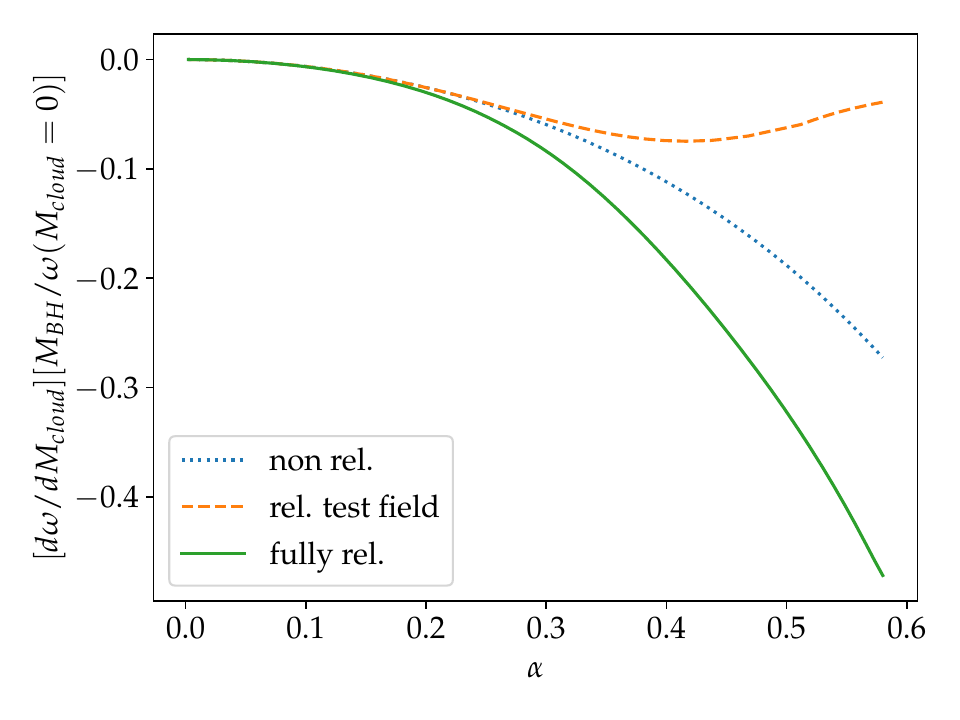} \label{fig:shiftValpha_b}}}%
    \caption{The cloud mass dependence of the frequency of oscillation of the cloud around the black hole for scalar bosons (top) and vector bosons (bottom). The leading order nonrelativistic result ($\propto \alpha^3$) is plotted along with the relativistic test field result from Ref.~\cite{Siemonsen:2022yyf} and the fully relativistic result of this paper.
    }%
    \label{fig:shiftValpha}
\end{figure}

We begin by comparing the relativistic frequency shift for scalar and vector boson clouds calculated in this work
to the nonrelativistic approximation described in Sec.~\ref{sec:non-rel_shift_calc}. 
In Fig.~\ref{fig:shiftValpha}, we plot the quantity $d\omega/dM_{\text{cloud}}$ at $M_{\text{cloud}} = 0$,
which means taking only
the part of the frequency shift linear in cloud mass and ignoring any higher
order parts. (As shown in Appendix~\ref{sec:higher_order_mc_fit}, the quadratic in cloud mass term
in the relativistic case contributes at the percent level.) 
From the plot, it is apparent that the deviation of the relativistic calculation from the nonrelativistic limit can be
as large as 
$40\%-50\%$ in the scalar case and $70\%-80\%$ in the vector case. 

In Fig.~\ref{fig:shiftValpha}, we also compare our results to the
quasirelativistic calculation of the frequency shift used in Ref.
\cite{Siemonsen:2022yyf}, which is computed using the energy density computed
from the relativistic test field, but still utilizing the Newtonian expression
for the frequency shift in terms of this quantity. We find that the test field
calculation does well in the scalar case up to moderately high values of
$\alpha$, significantly improving over the nonrelativistic expression up until
$\alpha\approx 0.35$.  However, for $\alpha > 0.37$ the test field calculation
actually becomes less accurate than the leading order nonrelativistic limit\footnote{In Fig. 6 of Ref.~\cite{Siemonsen:2022yyf}, there is a plotting error in the test field calculation shown due to the use of an expression for the black hole spin that is inaccurate for high spin or high $\alpha$.}. 
In the vector case, the test field calculation always performs worse than
the nonrelativistic limit. 
It seems that in this case including only some relativistic corrections can actually
give a worse result than the purely nonrelativistic one, though there may
be a more accurate way to generalize the Newtonian expression for the frequency
shift to the relativistic regime. 

We next comment on the estimated errors in the fully relativistic calculation of the frequency shift.
As described in Appendix \ref{app:errors}, we use a combination of numerical error due to finite resolution, theoretical error due to our oscillation-averaged approximation, and interpolation error due to the finite number of numerical points to estimate a total error in the evaluation of the frequency shift. 
A summary of these different error estimates can be found 
at the beginning of Appendix \ref{app:errors}.
As can be seen there, in the scalar case, the relative error in the frequency shift ranges from a few times $0.1\%$ to $1\%$ across
the parameter space, and is dominated by the interpolation error. For the vector case,
on the other hand, the theoretical error is larger, and dominates over the numerical error and the interpolation error in much of
the parameter space. The relative error in the vector frequency shift ranges from $0.001\%$ at low $\alpha$ to $1\%$ at the highest values of $\alpha$ considered.

\subsection{Implications for gravitational wave searches for ultralight bosons}
\label{sec:implications_bosons}

\begin{figure}[h]
\centering
\includegraphics[width=8cm]{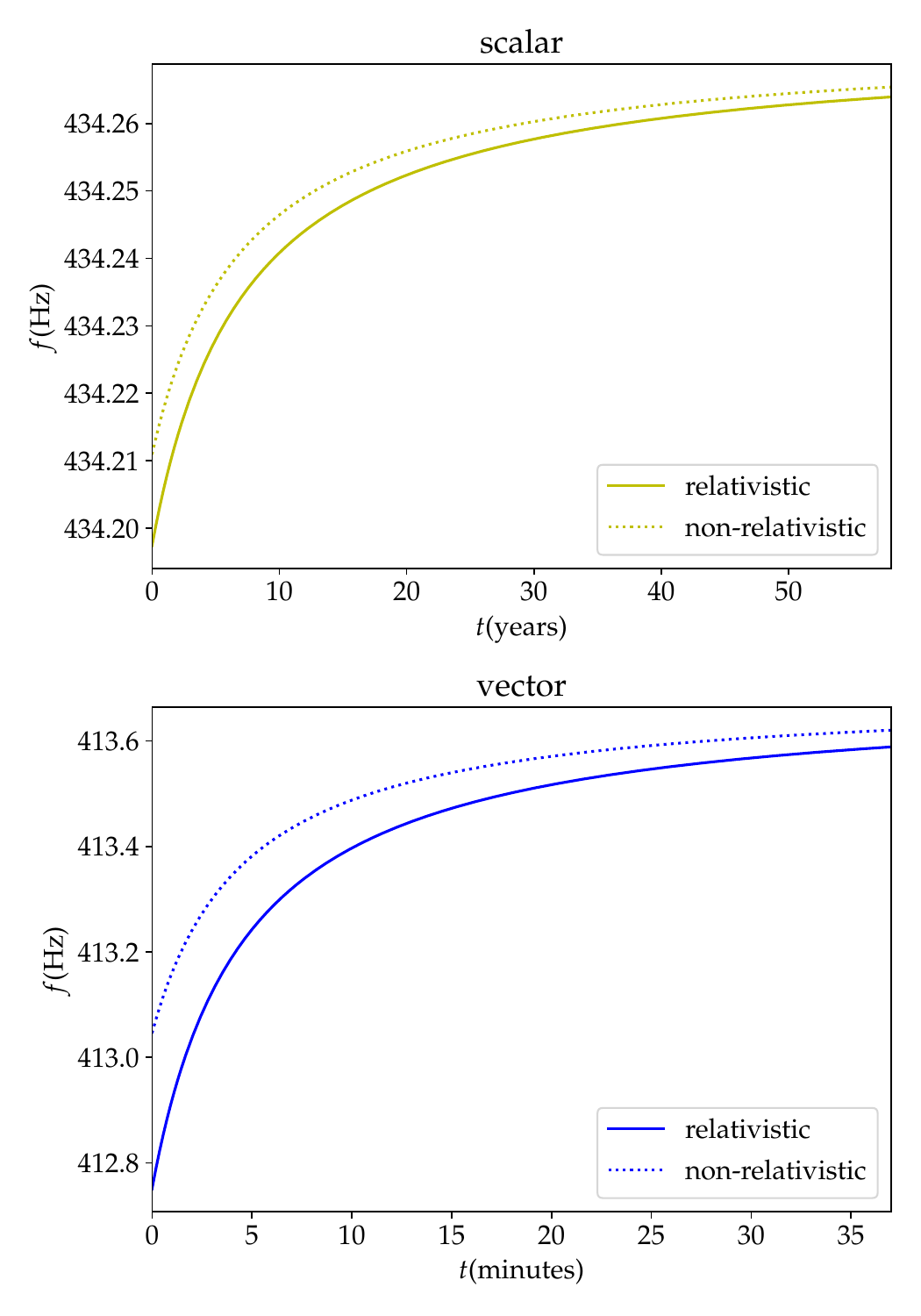}
\caption{The frequency evolution for an example waveform with an
initial black hole of dimensionless spin
$\chi_{\text{BH}} = 0.91$ and mass $M_{\text{BH}} = 47\ M_\odot$, and
an ultralight boson with mass $m_b = 9.1\times10^{-13}$ eV (corresponding to
$\alpha = 0.32$). Here $f$ is the gravitational wave frequency, plotted from the maximum cloud mass of $M_{\text{cloud}}/M_{\text{BH}}\approx 3\%$ ($5\%$) to the point where $M_{\text{cloud}}/M_{\text{BH}}\approx0.3\%$ $(0.5\%)$ for the
scalar (vector) case.
}
\label{fig:waveform_example}
\end{figure}

The more accurate relativistic frequency shift calculations described in the
previous section can be used to better model the frequency evolution of the
gravitational wave signal from an oscillating boson cloud.  We have
incorporated these results into \texttt{SuperRad}, a waveform model for
superradiant clouds described in Ref.~\cite{Siemonsen:2022yyf}. In Fig.
\ref{fig:waveform_example}, we plot example waveforms for scalar and vector
boson cases for an initial black hole with properties consistent with the remnant from the binary black hole merger GW200308 \cite{KAGRA:2021vkt}.
Here we can see the enhanced frequency shift in the fully relativistic
calculation compared to the nonrelativistic estimate, corresponding to a larger change in
frequency and a stronger chirp as the cloud dissipates through
gravitational radiation.

We roughly estimate the phase error over one gravitational wave emission
timescale with the relativistic calculation of the frequency shift, using the
estimated relative error $\delta\omega/\Delta\omega$, as described in Appendix
\ref{app:errors}. Recalling that the gravitational wave frequency
is given by twice the oscillation frequency of the cloud, the additional gravitational wave
phase offset due to the frequency shift is
\begin{align}
    \Delta \phi &= 2\int_0^{\tau_{\rm{GW}}}dt\ \left[\omega(t)-\omega(0)\right] 
    ,
\end{align}
where here the time $t=0$ corresponds the end of the superradiant instability phase and the
peak of the signal, before the cloud starts dissipating, and $\tau_{\rm{GW}}$ is the
gravitational wave emission timescale. We translate our estimate of the error in the frequency
shift into an estimate of the phase error in the updated waveform using
\begin{align}
    E(\Delta \phi) &= 2\int_0^{\tau_{\rm{GW}}}dt\ \delta\omega 
\approx \Delta \phi  \left ( \frac{\delta\omega}{\Delta \omega} \right)_{t=0}
    .
\end{align}

\begin{figure}
    \centering   
    \subfloat[\centering Initial black hole with dimensionless spin $\chi_{\text{BH}} = 0.99$.]{{\includegraphics[width=8cm]{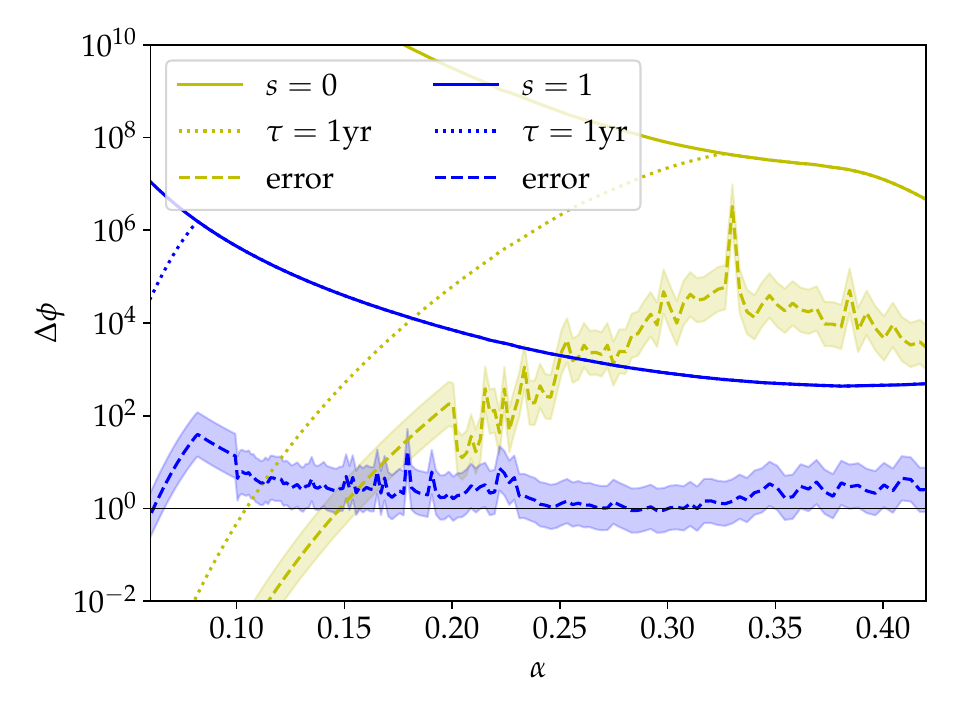} \label{fig:phase_offset_a}}}%
    \qquad
    \subfloat[\centering Initial black hole with dimensionless spin $\chi_{\text{BH}} = 0.7$.]{{\includegraphics[width=8cm]{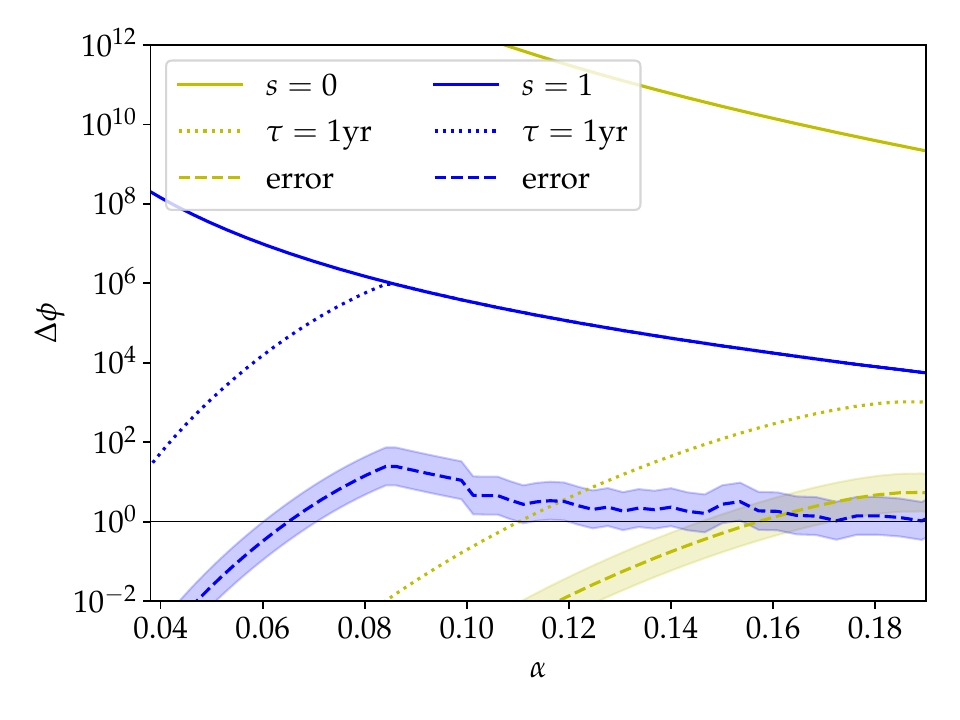} \label{fig:phase_offset_b}}}%
    \caption{The phase offset after integrating the gravitational wave signal over its emission timescale, or one year if the gravitational wave timescale is longer than that. In each case the initial black hole mass is $M_{\text{BH}} = 50\ M_\odot$. The jump in error at low $\alpha$ is where the relativistic shift calculation is extrapolated. The shaded region around the error in phase offset corresponds to an uncertainty in the error of a factor of three, indicating that the error shown here is an order of magnitude estimate.
    }%
    \label{fig:phase_offset}
\end{figure}

The accumulated phase offset $\Delta \phi$ for scalar and vector clouds
with different values of $\alpha$, as well as the estimated error in this
quantity, is shown in Fig. \ref{fig:phase_offset}. We include both a case where the
initial black hole is rapidly spinning, to illustrate the behavior at high
values of $\alpha$, as well as a case with a more moderate spin typical of the
remnant of a comparable mass binary black hole merger. The figure illustrates
that, except for scalar bosons and low values $\alpha$, where the gravitational
wave signal does not evolve significantly over the maximum one year period
considered, this phase offset is significant, and the changing frequency needs to
be taken into account.  For the relativistically correct frequency shift
calculated here, we estimate that the theoretical error in much of the
parameter space for vector bosons is on the order of a few radians over the
characteristic timescale of the gravitational wave signal.  This is nearing the
regime where theoretical modelling errors would no longer be an impediment to
coherent gravitational wave searches, although there are other reasons that
such searches may not be practical, including uncertainty in black hole
parameters and computational expense. 

The estimated phase error in the scalar case is much larger compared to the vector case
over much of the parameter space, even when the relativistically correct frequency shift is
included. This is because of the longer gravitational wave emission timescale in the
scalar case. While the relative error in the frequency shift is
comparable to the vector case, when integrated over a much longer period of
time it leads to a larger phase error.

\begin{figure}[h]
\centering
\includegraphics[width=8cm]{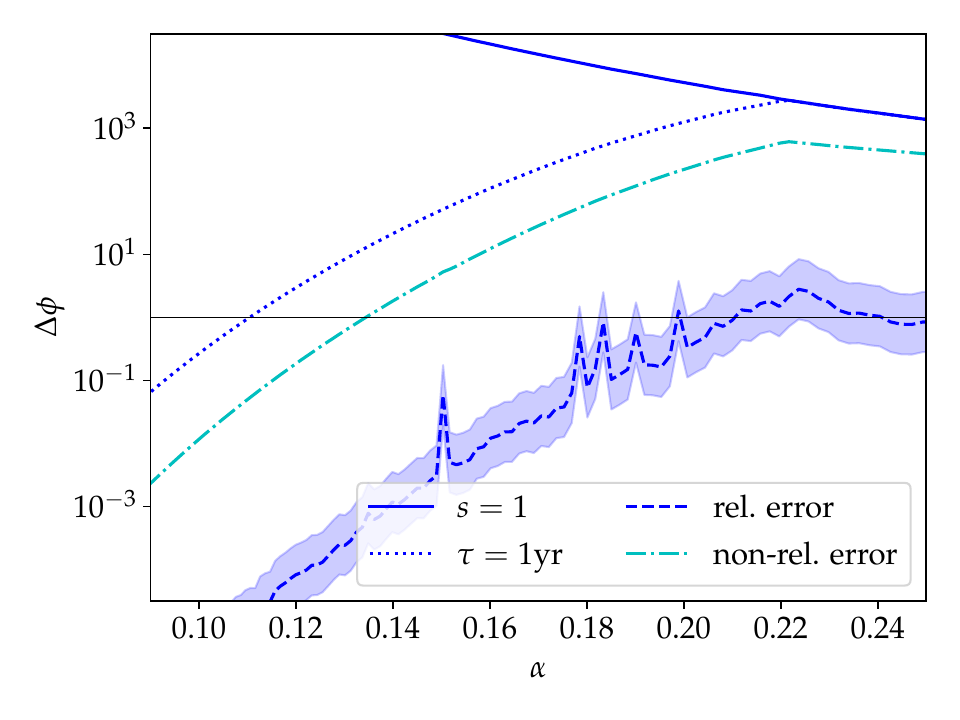}
\caption{The phase offset after integrating the gravitational wave signal over its emission timescale, or one year if the gravitational wave timescale is longer than that. Here we consider a LISA source, with initial black hole mass $M_{\text{BH}} = 5\times 10^{5}M_\odot\ $ and initial dimensionless spin $\chi = 0.85$. We compare the total accumulated phase, the error in the relativistic waveform, and the error in a waveform using the leading order nonrelativistic frequency shift.}
\label{fig:lisa_phase_offset}
\end{figure}

In addition to the ground based detector sources discussed above, we consider
an example relevant for LISA in Fig. \ref{fig:lisa_phase_offset}. 
Constituents of massive black hole binaries are expected to have higher dimensionless spins than 
stellar-mass black holes, and may also be promising candidates for binary follow-up searches for gravitational wave
from vector clouds around merger remnants \cite{Siemonsen:2022yyf}.
Following Ref. \cite{Lindblom:2008cm}, when the phase error is the dominant
source of error, a model with a maximum phase error of order
$1/\rho_{\text{max.}}$ is sufficiently accurate for a signal with maximum signal-to-noise ratio (SNR)
of $\rho_{\text{max.}}$. In Fig. \ref{fig:lisa_phase_offset}, we plot the
phase error for the vector case using two different models. One model uses the relativistic
frequency shift calculated in this work, and other uses the leading order
nonrelativistic frequency shift. 
%We mark with a black line 
The phase error corresponding to sufficiently accurate model for a signal with $\rho_{\text{max.}}=10$ is approximately $1/\rho_{\text{max.}} = 0.1$.
From the plot, it can be seen that while the nonrelativistic shift model is
sufficient for $\alpha<0.115$ when $\rho_{\text{max.}}=10$, the relativistic shift
model covers the larger range of $\alpha<0.18$. 
 It is evident
that the improved waveform model is preferred for $0.115<\alpha<0.18$ or for higher SNR
signals at lower $\alpha$.

\subsection{Geometry}
\label{sec:geometry}
In addition to changing the oscillation frequency, the boson cloud will change the spacetime geometry
around the black hole, which we can characterize using the solutions we construct. 
Here, we focus on calculating the change in the light ring and the ISCO of the black hole in the presence of a boson cloud, as these are related to how observations
of accreting black holes are used to measure the black hole spin. 
There has been significant work on
this topic in the past. A full null ray-tracing analysis was carried out for black holes with vector
boson clouds in Ref. \cite{Sengo:2022jif}, ISCO frequencies were calculated
for black holes with scalar hair in Ref. \cite{Herdeiro:2015gia}, and predictions
for $\text{K}\alpha$ emission lines have been calculated using null-ray-tracing
in Ref. \cite{Ni:2016rhz} for black holes with scalar hair. Here, we calculate equatorial null and test particle orbits around black holes
with both scalar and vector clouds. 

\begin{figure}
    \centering   
    \subfloat[\centering Scalar field]{{\includegraphics[width=8cm]{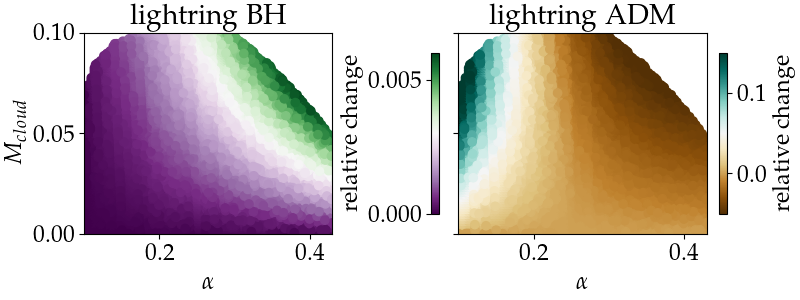} \label{fig:scalar_lr_isco}}}%
    \qquad
    \subfloat[\centering Vector field]{{\includegraphics[width=8cm]{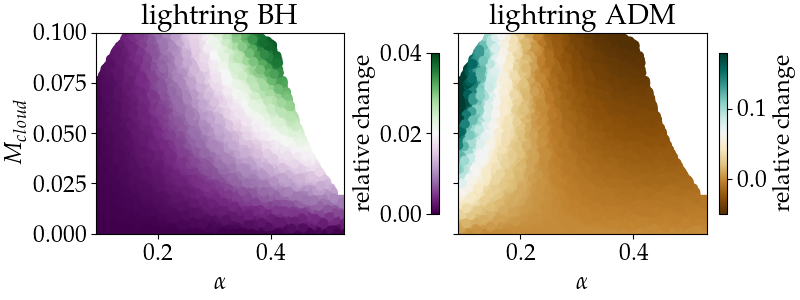} \label{fig:proca_lr_isco}}}%
    \caption{The relative difference in the circumferential radius of the light ring for a black hole with a boson cloud compared
to a Kerr black hole with the same black hole mass and spin (left panels) or the same global mass and angular momentum (right panels).
}%
    \label{fig:lr_isco}
\end{figure}

\begin{figure}
    \centering   
    \subfloat[\centering Scalar field]{{\includegraphics[width=8cm]{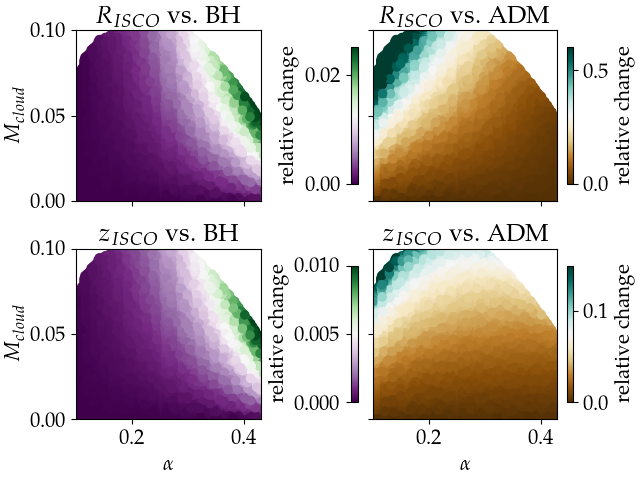} \label{fig:scalar_redshift}}}%
    \qquad
    \subfloat[\centering Vector field]{{\includegraphics[width=8cm]{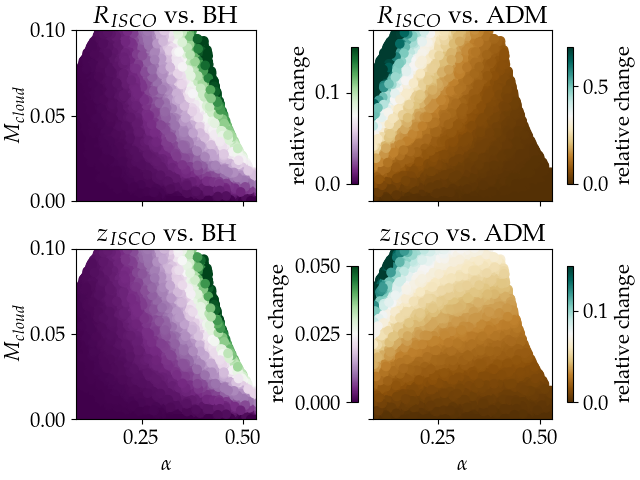} \label{fig:proca_redshift}}}%
    \caption{Here we show the ISCO radius and the redshift from the ISCO to
infinity in the presence of a boson cloud, compared to the ISCO radius and
redshift from the ISCO of a Kerr black hole with either the same central black
hole's properties or the same global mass and angular momentum. The quantity plotted is
the relative difference in $\omega_1$ and $\omega_2$ where $\omega_1$ is the
frequency measured by the observer traveling along the ISCO and $\omega_2$ is
the frequency measured by a stationary observer at infinite distance. The
result plotted here is for light emitted tangent to the observer's motion on
the ISCO and in the same (prograde) direction.
    }%
    \label{fig:redshift}
\end{figure}

For all the black hole-cloud solutions we construct, we calculate the position of the
equatorial light ring and ISCO around a black hole in the presence of a
superradiant cloud as is described in Appendix \ref{sec:appendix_lr_isco},
using the circumferential radius to record our measured radii in a coordinate
independent way. We compare the light ring and ISCO radii of a black hole with
a scalar or vector cloud with those of a Kerr black hole with either the same
black hole parameters (measured using the horizon quantities) or the same total 
mass and angular momentum (in particular we use the ADM mass~\cite{Arnowitt:1959ah}). 
The results for the deviation of the light ring from the Kerr values are shown in Fig. \ref{fig:lr_isco}
and the results for the ISCO radius are shown in Fig.~\ref{fig:redshift}.
As one might expect, the biggest effect comes from the comparison where the global quantities are kept
fixed, due to the significant amount of mass and angular momentum that can be carried in the cloud.
However, for the vector case at high $\alpha$, the cloud can have non-negligible effect on the near horizon geometry, shifting
the light ring and ISCO radius by up to a few percent. 

In addition to calculating the location of the ISCO for the black hole-boson
cloud systems, we calculate the ISCO frequency, and the redshift of light
emitted by an observer traveling along the ISCO as measured by another observer
at rest at infinity. In particular, we calculate the redshift for  a light ray
emitted tangent to the observer's trajectory in the same direction (prograde to
the black hole's spin), which we call $z_{\rm ISCO}$, and show the deviation in this quantity from its
Kerr value in Fig. \ref{fig:redshift}, either fixing the black hole or
the global mass and angular momentum.  This shows similar behavior to the ISCO
radius. We comment on how the deviations in these quantities from their Kerr
values might be related to spin mismeasurement in the presence of a boson
cloud in the next section.

The numerical values for the light ring radius, ISCO radius, frequency, and
redshift (for light emitted both in the same or opposite direction as the
emitter orbiting at the ISCO) of black holes with scalar and vector boson clouds are publicly available\footnote{
Data files with the geometric quantities and usage examples can found at \href{https://www.bitbucket.org/weast/superrad/}{https://www.bitbucket.org/weast/superrad}. }.

\subsection{Implications for Kerr parameter measurements}
\label{sec:implications_kerrparams}

One method for inferring black hole spins is to measure how the emission  
from accreting black holes is redshifted (or blue shifted)~\cite{McClintock:2013vwa,Reynolds:2019uxi,Reynolds:2013qqa}.
Roughly speaking, the maximum redshift is taken to come from the ISCO since any material
inside the ISCO is not expected to contribute to the observed electromagnetic spectrum 
under the assumption that the accretion disk is thin and oriented on
the equator. Since the presence of a cloud can affect the position and redshift at the ISCO,
we wish to estimate how much it would affect these spin measurements. 

There are two main methods for measuring black hole spins from spectra,
$\text{K}\alpha$ line broadening and continuum fitting. The $\text{K}\alpha$ line
broadening method yields a dimensionless redshift measurement that can be used to directly
fit the dimensionless spin \cite{Reynolds:2019uxi}. The continuum fitting method gives a
dimensionful ISCO radius measurement that must then be scaled by an independent
mass measurement, often from distant orbiting
dynamics~\cite{McClintock:2013vwa}. A mass measurement using distant orbiting
dynamics would give a value close to the total mass (e.g. in a black
hole-cloud system), rather than the central black hole mass.

\begin{figure}
    \centering   
    \subfloat[\centering Scalar field]{{\includegraphics[width=8cm]{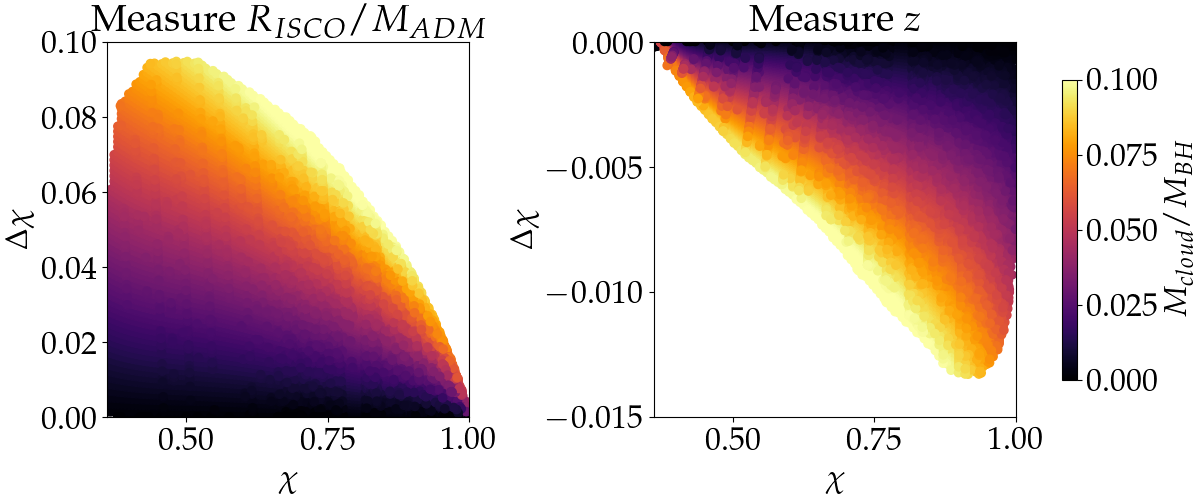} \label{fig:scalar_spinest}}}%
    \qquad
    \subfloat[\centering Vector field]{{\includegraphics[width=8cm]{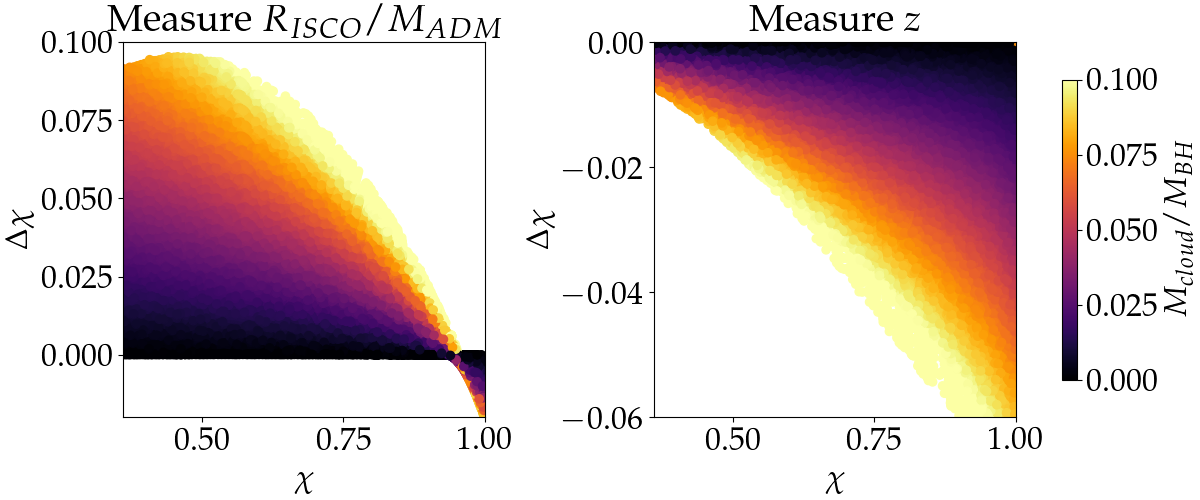} \label{fig:proca_spinest}}}%
    \caption{The effect of the presence of a boson cloud on the dimensionless spin measurements using a known mass and redshift measurements. The results shown are only for clouds that could be formed through superradiance around an isolated black hole. The spin shown here is the spin of the central black hole after the cloud's formation.
    }%
    \label{fig:spin_measurement_implications}
\end{figure}

A complete analysis of how the presence of a boson cloud would affect an accreting 
black hole would require solving for accretion
disk profiles on the black hole-cloud solutions and tracing out the emission that they
source, which is beyond the scope this work. 
We further note that we here we are restricting to our axisymmetric and stationary spacetimes,
and neglecting the effect of an oscillating component to the gravitational potential that
would be present in the case of single (real) scalar or vector field (see, e.g., Ref.~\cite{Chen:2022kzv}).
However, as a rough
estimate of the discrepancy between the dimensionless spin $\chi=J_{\rm BH}/M_{\rm BH}^2$ one would infer from
such methods when assuming a Kerr black hole when a boson cloud is present, 
we consider how the value of $z_{\rm ISCO}$, or the ratio $R_{\text{ISCO}}/M_{\text{ADM}}$,
maps to a dimensionless black hole spin assuming Kerr.
We define the difference between this 
estimated black hole spin $ \chi_{\text{est.}}$ and the true dimensionless spin
$\chi$ of a black hole surrounded by a boson cloud as
\begin{align}
    \Delta \chi = \chi_{\text{est.}}-\chi 
    .
\end{align}
We plot $\Delta \chi$ in Fig. \ref{fig:spin_measurement_implications} for the
both the case where one combines the dimensionful ISCO radius with the global mass (left panels),
and the case based on the dimensionless redshift $z_{\rm ISCO}$ (right panels). 
When considering $R_{\text{ISCO}}/M_{\text{ADM}}$, both the scalar and boson case show 
a similar overestimate in the dimensionless spin up to moderately high values, that
is mostly just due to the overestimate of the black hole mass. Interestingly, 
for the most relativistic vector boson cases, the effect on the geometry near
the black hole becomes significant, and causes this quantity to give an underestimate
of the black hole spin. Using the redshift at the ISCO has a smaller effect,
leading to a maximum overestimate in the dimensionless spin of $\sim1\%$ in the scalar case,
and $5\%$ in the vector case.

\section{Discussion and Conclusion}
\label{sec:discussion}

In this work, we have used fully general relativistic solutions to study
the self-gravity effects of boson clouds that might form around black holes
through the superradiant instability. In particular, we have focused on 
how the frequency of the cloud's oscillation changes due to the nonvanishing
mass of the cloud. This impacts the frequency shift and evolution of the gravitational
waves emitted by the clouds. We found that relativistic effects can enhance this frequency
shift by up to a factor of two over nonrelativistic estimates obtained
previously. Further, our current fully relativistic computation of the frequency shift 
is accurate up to uncertainties at the percent level or better. The remaining errors are due 
both to finite numerical resolution, as well as theoretical error arising
from an oscillation-average approximation taken here.

Using this fully relativistic frequency shift, we improved the gravitational waveform
model \texttt{SuperRad}~\cite{Siemonsen:2022yyf}. This allows for
accurate modeling of the loudest, and thus fastest evolving, boson cloud
gravitational wave signals (i.e., those where the change in frequency during 
the observation time is most significant). This is particularly relevant
for targeting gravitational wave emitted by ultralight vector boson clouds 
around binary black hole remnants. Currently,
proposed searches for such signals are based on semicoherent search methods, which
are fast and more robust to modeling errors, at the expense of have a
reduced sensitivity compared to a fully coherent searches~\cite{Jones:2023fzz}.
The results here can be used to better tune the parameters of such searches,
but may also be leveraged to design more sensitive search techniques. 
This is relevant not only for current ground-based gravitational
wave detectors, but also for supermassive black hole mergers and
space-based detectors such as LISA.

Most previous continuous wave searches for signals from boson clouds have
focused on scalar bosons in the small $\alpha$ regime (for example, in
Ref.~\cite{LIGOScientific:2021rnv} the largest value of $\alpha$ considered is
$\alpha = 0.15$) and often use frequency binning methods that assume the
frequency variation within a timestep (or a minimum coherence time) is below
a given threshold. In these cases, the relativistic correction to the rate of frequency
evolution, and therefore the correction to the maximum $\alpha$ constrained, 
is at the level of only a few percent. However, for larger values of $\alpha$, this correction will be
more significant. For example, Ref.~\cite{Zhu:2020tht} estimated the
detectability of scalar boson clouds with $\alpha \leq 0.5$ for blind searches,
and Ref.~\cite{Jones:2023fzz} estimated the detectability of targeted searches
for vector boson clouds with $\alpha\lesssim 0.2$. Considering the typical
remnant of a comparable mass, low-spin binary black hole merger, the strain is
maximized for $\alpha \approx 0.17$, where the relativistic frequency shift
correction is $\approx 10\%$ in the vector boson case.  

In this work, we have not considered bosonic fields with nongravitational
self-interactions or interactions with other fields, which may give rise to
additional radiation channels or affect the saturation the superradiant
instability and the expected gravitational wave signal
\cite{Fukuda:2019ewf,Baryakhtar:2020gao,East:2022ppo,East:2022rsi,Chia:2022udn,Siemonsen:2022ivj,Omiya:2022gwu,Omiya:2024xlz,Collaviti:2024mvh}.
For future work, it would be interesting to see whether relativistic black
hole-boson cloud solutions like those considered here can be used to incorporate 
the impact of these interactions on the observational signatures (see, e.g., Ref.~\cite{Herdeiro:2015tia}).

Finally, we have also studied how the boson cloud will change the black hole spacetime,
and commented on how this may affect spin measurements based on observations of
accreting black holes. As is discussed in Ref. \cite{Khalaf:2024nwc}, since the
two main methods of spin measurement from redshifts, continuum fitting and
$\text{K}\alpha$ line broadening, give different biases in the presence of a
boson cloud, the difference between the values of these two measurements can be
used to probe the existence of extra matter located close to the black hole.
As was assumed in Ref.~\cite{Khalaf:2024nwc}, we find here that the discrepancy
in the spin from mapping the dimensionless redshift to its value in an isolated
black hole is subdominant to the discrepancy due to the use of the dimensionful
ISCO measurement in the scalar case. In the vector case, however, the former
can be more significant at moderately high spin due to the stronger effect on
the geometry in the vicinity of the black hole horizon, potentially giving rise
to a more sensitive probe for the presence of a boson cloud. 

\section{Acknowledgments}

The authors would like to thank Dana Jones and Ling Sun for helpful discussions.
W.E. and T.M. acknowledge support from a Natural Sciences and Engineering Research
Council of Canada Discovery Grant and an Ontario Ministry of Colleges and
Universities Early Researcher Award. This research was supported in part by
Perimeter Institute for Theoretical Physics. Research at Perimeter Institute is
supported in part by the Government of Canada through the Department of
Innovation, Science and Economic Development and by the Province of Ontario
through the Ministry of Colleges and Universities. 
This research was enabled in part by support
provided by SciNet (www.scinethpc.ca) and the
Digital Research Alliance of Canada (alliancecan.ca). Calculations were
performed on the Symmetry cluster at Perimeter Institute, 
the Niagara cluster at the University of 
Toronto, and the Narval cluster at Ecole de technologie sup\'erieure 
in Montreal.

\section{Data availability}

The data that support the findings of this article are openly available \cite{dataref}.

\appendix

\section{Physical quantities}
\label{sec:physical_quantities}
For the black hole-boson cloud solutions constructed in this work, we calculate several quantities that characterize the mass and angular momentum associated with the black hole and the cloud.
To calculate the boson cloud mass, we use the Komar mass integral, as is done in Refs.~\cite{Herdeiro:2015gia,Herdeiro:2016tmi,Santos:2020pmh}, with
\begin{align}
\label{eq:cloud_mass_int}
    M_{\text{cloud}} = -\int_\Sigma dS_a\left(2T^a_b t^b-T t^a\right)
    ,
\end{align}
where $T_{ab}$ is the stress energy tensor with trace $T$,
the integral is over a spacelike hypersurface $\Sigma$, with $dS_a = n_a dV$ proportional to the unit future pointing normal $n_a$ to that hypersurface, and $t^b$ is the Killing vector associated with the stationary spacetime.
The cloud angular momentum $J_{\rm cloud}$ is calculated using the Komar
integral as in the cloud mass integral in Eq.~\eqref{eq:cloud_mass_int}, but with
$t^a$ replaced by the azimuthal Killing vector $\varphi^a$. For both these
quantities, we use a special treatment of the inner boundary, to avoid error
due to setting the inner radial boundary slightly off the black hole horizon.
This involves interpolating the integrand over the last few points before the
horizon point then extrapolating to the point on the horizon. We use a linear
interpolation because the extrapolated point converges well in that case.

We calculate the black hole mass using two different approaches. We can either
calculate the ADM mass and subtract the cloud mass or directly calculate the
horizon mass. In different regimes, these two calculations have better
associated error and so we calculate both in order to minimize error as well as
to check our mass and angular momentum calculations are consistent.
The ADM mass and total angular momentum $J$ can be calculated using the asymptotic behavior of the metric as is described in Ref.~\cite{Herdeiro:2015gia} (Eq. 3.15). We also calculate the horizon mass as is described in Ref.~\cite{Herdeiro:2015gia} using the expression
\begin{align}
    M_{\text{BH}} = 2T_HS+2\Omega_H (J-J_{\text{cloud}})
    ,
\end{align}
where the Hawking temperature $T_H$, and the entropy $S$ are calculated according to Ref.~\cite{Herdeiro:2015gia}. 

In the scalar case, we use the horizon mass and the Komar cloud mass in all of
the parameter space and in the vector case, we use the ADM mass and the Komar
cloud mass for high $\alpha$ and the horizon mass and the Komar cloud mass for
lower $\alpha$. In each case, we check that these two definitions give results
that are consistent within the numerical error.

\section{Constructing solutions}
\label{sec:eom}

\subsection{Elliptic differential equations}
\label{sec:listing_diff_equations}

We formulate differential equations using the Einstein-Klein Gordon and Einstein-Proca equations (along with a Lorenz condition in the Proca case).
The solution is then constructed from an ansatz. We use metric ansatz with an explicit horizon in the Lewis–Papapetrou coordinates
\begin{align}\label{metric_ansatz}
    ds^2 &= -fNdt^2 +\frac{l}{f}\left[g\left(\frac{dr^2}{N}+r^2d\theta^2\right)\right.\nonumber\\
    &\left.+r^2\sin^2\theta \left(d\varphi - \frac{\omega_m}{r}dt\right)^2\right]
,
\end{align}
where $N = (r-r_H)/r$ vanishes at the horizon radius. This ansatz is stationary
and axisymmetric, and so the metric functions $f$, $l$, $g$, and  $\omega_m$ are real
valued functions of only two coordinates ($r$ and $\theta$) and can be solved
on a two dimensional grid. Due to the equatorial symmetry, we can restrict the
angular domain to the region above the equatorial plane, $\theta \in (0,\pi/2)$. We use a compactified radial
coordinate $x = (r-r_H)/r$, and include slight offsets from the
boundaries of $0<x<1$ and $0<\theta<\pi/2$ such that our numerical grid
boundaries are at $10^{-7}<x<1-10^{-7}$ and $10^{-7}<\theta<\pi/2-10^{-7}$. The
matter field values are solved simultaneously on the same grid. The scalar
field ansatz is
\begin{align}\label{complex_scalar_ansatz}
    \Phi(t,r,\theta, \varphi) &= \phi(r,\theta) e^{-i\omega t + i m \varphi} .
\end{align}
Here $\phi(r,\theta)$ is a real valued function, $m \in \mathbb{Z}$, and $\omega \in \mathbb{R}$. The function $\phi$ is solved along with the metric functions in $r$-$\theta$ space to give a full solution. In the vector case we use the ansatz
\begin{align}\label{complex_proca_ansatz}
    A_0(t,r,\theta, \varphi) &= i V(r,\theta) e^{-i\omega t + i m \varphi}\\
    A_r(t,r,\theta, \varphi) &= H_1(r,\theta) e^{-i\omega t + i m \varphi}\\
    A_\theta(t,r,\theta, \varphi) &= H_2(r,\theta) e^{-i\omega t + i m \varphi}\\
    A_\varphi(t,r,\theta, \varphi) &= i \sin(\theta) H_3(r,\theta) e^{-i\omega t + i m \varphi}
\end{align}
and solve for real valued functions $V(r,\theta)$, $H_1(r,\theta)$, $H_2(r,\theta)$, and $H_3(r,\theta)$.
The choice of $\omega$ and $r_H$ (in units of $\mu$) determines the solution.

We formulate the scalar (vector) equations of motion by combining Einstein's equation with the Klein Gordon (Proca and Lorentz) equation(s). We get a set of five (eight) second order differential equations, such that we have
\begin{align}
    \partial^2F_i = \text{Polynomial}(\partial F_j, F_j)
\end{align}
for each real valued field $F_i$. 
We use a metric ansatz that is different from
Refs.~\cite{Herdeiro:2014goa,Herdeiro:2015gia,Herdeiro:2016tmi,Santos:2020pmh},
and we find that, while this ansatz is well behaved in the boson star case,
there is a numerical divergence at the horizon in the black hole with boson
cloud case. We remove that divergence by including an overall factor of $x^3$
in all the equations of motion given below. 
We list the equations of motion for a complex vector field in the presence of a black hole here,
these expressions are also available in a notebook accompanying this article.
They are 
\begin{widetext}
\begin{align}
    \partial_r^2 V &+ \partial_\theta^2 V /\left(r (r-r_H)\right) = -\left((r_H-r) \sin (\theta ) \left(-\csc (\theta ) \partial_r l \partial_r V r^3-2 \omega_m \partial_r H_3 \partial_r l r^2-2 l \partial_r H_3 \partial_r \omega_m r^2-4 \csc (\theta ) l \partial_r V r^2\right.\right.\nonumber\\
    &+r_H \csc (\theta ) \partial_r l \partial_r V r^2+2 m^2 \csc ^3(\theta ) g l V r-2 \cot (\theta ) \csc (\theta ) l \partial_\theta V r-\csc (\theta )
   \partial_\theta l \partial_\theta V r-2 l \omega_m \partial_r H_3 r+2 r_H
   \omega_m \partial_r H_3 \partial_r l r
   \nonumber\\
   &+2 r_H l \partial_r H_3 \partial_r \omega_m r+4 r_H \csc (\theta )
   l \partial_r V r-4 \cot ^2(\theta ) H_3 l \omega_m-4 \cot (\theta ) l \omega_m
   \partial_\theta H_3-2 \cot (\theta ) H_3 \omega_m \partial_\theta l
   \nonumber\\
   &-2 \omega_m \partial_\theta H_3 \partial_\theta l-2 \cot
   (\theta ) H_3 l \partial_\theta \omega_m-2 l \partial_\theta H_3 \partial_\theta \omega_m+2 m \csc (\theta ) H_2
   \left(\omega_m \partial_\theta l+l \left(2 \cot (\theta ) \omega_m+\partial_\theta \omega_m\right)\right) 
   \nonumber\\
   &\left.+4 r_H l
   \omega_m \partial_r H_3-2 \csc (\theta ) H_1 \left(m r (r_H-r) \omega_m \partial_r l+l \left(-r r_H
   \omega +m (2 r_H-r) \omega_m+m r (r_H-r) \partial_r \omega_m\right)\right)\right)\nonumber\\
   &f^2
   +2 (r_H-r) l \left(\mu^2
   g l V r^3+\omega  H_1 \partial_r f r^3+\partial_r f \partial_r V r^3-r_H \omega 
   H_1 \partial_r f r^2-2 m H_1 \omega_m \partial_r f r^2+2 \omega_m \sin (\theta ) \partial_r f
   \partial_r H_3 r^2
   \right.
   \nonumber\\
   &-r_H \partial_r f \partial_r V r^2+\partial_\theta f \partial_\theta V r+2 m r_H H_1 \omega_m
   \partial_r f r-2 r_H \omega_m \sin (\theta ) \partial_r f \partial_r H_3 r+2 \cos (\theta ) H_3 \omega_m
   \partial_\theta f-H_2 (-r \omega +2 m \omega_m) \partial_\theta f
   \nonumber\\
   &\left.+2 \omega_m \sin (\theta ) \partial_\theta f
   \partial_\theta H_3\right) f+2 r l^2 \left(r g V (-r \omega +m \omega_m)^2-(r_H-r) H_1
   \omega_m \sin ^2(\theta ) \left(\omega_m-r \partial_r \omega_m\right) (-r \omega +m \omega_m)\right.
   \nonumber\\
   &+\omega_m \sin ^2(\theta )
   \left(\partial_r \omega_m \partial_r V r^3+\omega_m \sin (\theta ) \partial_r H_3 \partial_r \omega_m r^2-\omega_m \partial_r V
   r^2-r_H \partial_r \omega_m \partial_r V r^2+\partial_\theta \omega_m \partial_\theta V r-\omega_m^2 \sin (\theta ) \partial_r H_3 r
   \right.
   \nonumber\\
   &-r_H
   \omega_m \sin (\theta ) \partial_r H_3 \partial_r \omega_m r+r_H \omega_m \partial_r V r+\cos (\theta ) H_3 \omega_m \partial_\theta \omega_m-H_2 (-r \omega +m \omega_m) \partial_\theta \omega_m+\omega_m \sin (\theta ) \partial_\theta H_3
   \nonumber\\
   &\left.\left.\left.\partial_\theta \omega_m
   +r_H \omega_m^2 \sin (\theta ) \partial_r H_3\right)\right)\right)\left(2 r^2 (r_H-r)^2 f^2 l\right)^{-1}
   %\\
    ,
\end{align}
\begin{align}
    \partial_r^2H_1 &+\frac{ 
   \partial_\theta^2H_1}{r (r-r_H)}=\left(\frac{2 \cot (\theta) (r_H-r) \partial_\theta H_1}{r}-\frac{4 (r_H-r) \partial_\theta H_2}{r^2} \right.   \nonumber
   \\
   &+\frac{2 g l (m \omega_m-\omega  r) (H_3 \omega_m \sin (\theta
   +r V(r,\theta)) \partial_r f}{f^3}+4 (r_H-r) \partial_r H_1-\frac{2 (r_H-r) \partial_\theta g
   \left(\partial_\theta H_1-\partial_r H_2\right)}{g r}\nonumber
   \\
   &-\frac{(r_H-r) \partial_\theta l \left(\partial_\theta H_1-\partial_r H_2\right)}{l r}+\frac{(r_H-r) \partial_\theta l \partial_r H_2}{l r}-\frac{2
   (r_H-r) \left(\mu^2 g H_1 l r^2+\partial_\theta f \left(\partial_r H_2-\partial_\theta H_1\right)\right)}{f r}\nonumber
   \\
   &-\frac{2 m \csc (\theta) g (r_H-r) \left(m \csc (\theta) H_1-\partial_r H_3\right)}{r}+\frac{(r_H-r) \left((r_H-r) r \partial_r H_1-2 \partial_\theta H_2\right) \partial_r l}{l r}\nonumber
   \\
   &+\frac{2 m \csc (\theta) g (r_H-r) \left(H_3 \left(2 l+r \partial_r l\right)-l r \partial_r H_3\right)}{l r^2}-\left((r_H-r) \partial_r f \left(-H_2
   \left(2 \cot (\theta) l+\partial_\theta l\right)+2 l \left(m \csc (\theta) g H_3 \right.\right. \right.\nonumber
   \\
   &\left.\left.\left.-\partial_\theta H_2+(r_H-r) r \partial_r H_1\right)+H_1 \left(2 l (r_H-2 r)+(r_H-r) r
   \partial_r l\right)\right)\right)(f l r)^{-1}\nonumber
   \\
   &+\frac{(r_H-r) \partial_r g \left(-H_2 \left(2 \cot
   (\theta) l+\partial_\theta l\right)-2 l \left(\partial_\theta H_2+r (r-r_H) \partial_rH_1\right)+H_1 \left(2 l (r_H-2 r)+(r_H-r) r \partial_r l\right)\right)}{g l
   r}\nonumber
   \\
   &-\frac{2 g l (m \omega_m-\omega  r) \left(H_1 (m \omega_m-\omega  r)-\omega_m
   \sin (\theta) \partial_r H_3-r \partial_r V\right)}{f^2}\nonumber
   \\
   &-\left(2 g l \left(H_3
   \sin (\theta) \left(m (r_H-2 r) \omega_m^2-r \left(-\omega  r+2 m (r_H-r) \partial_r \omega_m\right) \omega_m-\omega  r^2
   (r-r_H) \partial_r \omega_m\right)\right.\right.\nonumber
   \\
   &\left.\left.+r \left(-r V(r,\theta) \left(-r_H \omega +m \omega_m+m (r_H-r) \partial_r\omega_m\right)-(r_H-r) (m \omega_m-\omega  r) \left(\omega_m \sin (\theta) \partial_r H_3+r \partial_r V\right)\right)\right)\right)\nonumber
   \\
   &(f^2 r (r-r_H))^{-1}-\frac{H_2 (r_H-r) \left(4 \cot (\theta) l^2+\left(2
   \partial_\theta l+2 \cot (\theta) r \partial_r l-r \partial_r\partial_\theta l\right) l+2 r \partial_\theta l
   \partial_r l\right)}{l^2 r^2}\nonumber
   \\
   &\left.+\frac{H_1 (r_H-r) \left(4 (r_H-r) l^2+r \left((3 r_H-4 r)
   \partial_r l+r (r-r_H) \partial_r^2l\right) l+2 (r_H-r) r^2 (\partial_r l)^2\right)}{l^2 r^2}\right)(2 (r_H-r)^2)^{-1}
   %\\
   %
    ,
\end{align}
\begin{align}
   \partial_r^2 H_2&=\left(\frac{2 (r_H-r) \left((r_H-r) \partial_r f \left(\partial_\theta H_1-\partial_r H_2\right)-\mu^2 r
   g H_2 l\right)}{f}\right.\nonumber\\
   &-\frac{2 g l (m \omega_m-r \omega ) \left(H_2
   (m \omega_m-r \omega )-\sin (\theta ) \partial_\theta H_3 \omega_m-\cos (\theta ) H_3 \omega_m-r \partial_\theta V\right)}{f^2}\nonumber\\
   &+\left((r_H-r) \left(-2 r (r_H-r) \partial_r g l \left(\partial_\theta H_1-\partial_r H_2\right)+g \left(-r (r_H-r) \partial_r l \left(\partial_\theta H_1-\partial_r H_2\right)-2 l \left(r_H
   \partial_\theta H_1\right.\right.\right.\right.\nonumber\\
   &\left.\left.\left.\left.\left.+r (r-r_H) \partial_r\partial_\theta H_1-r_H \partial_r H_2\right)\right)+2 m \csc ^2(\theta ) g^2 l \left(-m
   H_2+\sin (\theta ) \partial_\theta H_3+\cos (\theta ) H_3\right)\right)\right)(r g l)\right)(2 (r_H-r)^2)^{-1}
   %\\
\end{align}
\begin{align}
   \partial_r^2 H_3&+\partial_\theta^2 H_3/\left(r(r-r_H)\right)=\left(4 (r_H-r) f l \left(-m r^2 \csc (\theta ) \partial_r f H_1+m r r_H \csc (\theta )
   \partial_r f H_1-m \csc (\theta ) \partial_\theta f H_2+r^2 \partial_r f \partial_r H_3\right.\right. \nonumber
   \\
   &\left.+\partial_\theta f \partial_\theta H_3 -r r_H \partial_r f \partial_r H_3+\cot (\theta ) \partial_\theta f H_3-\mu^2 r^2 g H_3 l\right)-\left(\csc ^2(\theta ) (r_H-r) f^2 \left(H_3 \left(4 l
   \left(m^2 g+1\right)+\sin (2 \theta ) \partial_\theta l\right) \right.\right.\nonumber
   \\
   &-2 \sin (\theta ) \left(-2 m (r_H-r) H_1 \left(r \partial_r l+2
   l\right)-\sin (\theta ) \left(\partial_\theta H_3 \partial_\theta l+r (r-r_H) \partial_r H_3 \partial_r l\right)\right. \nonumber
   \\
   &\left.\left.\left.+2
   l \left(\cos (\theta ) \partial_\theta H_3+\sin (\theta ) \left(r_H \partial_r H_3\right)\right)\right)-4 m H_2 \left(\sin (\theta ) \partial_\theta l+2 \cos (\theta ) l\right)\right)\right)+2 r l^2 \left(2 \sin (\theta ) \left((r_H-r) H_1 \right.\right.\nonumber
   \\
   &
   \left(r \partial_r \omega_m-\omega_m\right) (m \omega_m-r \omega )-H_2 \partial_\theta \omega_m (m \omega_m-r \omega )+r^2 \sin (\theta )
   \partial_r H_3 \omega_m \partial_r \omega_m+\sin (\theta ) \partial_\theta H_3 \omega_m \partial_\theta \omega_m \nonumber
   \\
   &-r r_H \sin (\theta )
   \partial_r H_3 \omega_m \partial_r \omega_m-r \sin (\theta ) \partial_r H_3 \omega_m^2+r_H \sin (\theta ) \partial_r H_3
   \omega_m^2+\cos (\theta ) H_3 \omega_m \partial_\theta \omega_m+r^3 \partial_r \omega_m \partial_r V \nonumber
   \\
   &\left.\left.\left.-r^2 r_H \partial_r\omega_m \partial_r V+r \partial_\theta \omega_m \partial_\theta V-r^2 \omega_m \partial_r V+r r_H \omega_m \partial_r V\right)-2 g
   H_3 (m \omega_m-r \omega )^2\right)\right)(4 r (r_H-r)^2 f^2 l)^{-1}
   %\\
    ,
\end{align}
\begin{align}
    \partial_\theta^2&f+r(r-r_H)\partial_r^2f=\frac{1}{8} \left(\left(4 \kappa  \csc ^2(\theta ) \left(2 r (r-r_H) \sin ^2(\theta ) \left(\partial_\theta H_1-\partial_r H_2\right)^2+g \left(2 m^2 r
   (r-r_H) H_1^2-4 m r (r-r_H) \sin (\theta ) \partial_r H_3 H_1 \right.\right.\right.\right.\nonumber
   \\
   &+2 m^2 H_2^2+\cos (2 \theta ) H_3^2+H_3^2-\cos (2 \theta ) \partial_\theta H_3^2+\partial_\theta H_3^2+r^2 \partial_r H_3^2-r r_H \partial_r H_3^2-r^2 \cos (2 \theta )
   \partial_r H_3^2+r r_H \cos (2 \theta ) \partial_r H_3^2\nonumber
   \\
   &\left.\left.\left.+2 H_3 \sin (2 \theta ) \partial_\theta H_3-4 m H_2 \left(\cos (\theta
   ) H_3+\sin (\theta ) \partial_\theta H_3\right)\right)\right) f^2\right)(r^2 g l)^{-1}-\frac{4 r_H \partial_r l f}{l}+\left(2 \csc (\theta ) \left(4 \kappa  \sin (\theta ) \partial_r V^2 r^4\right.\right.\nonumber
   \\
   &-\frac{2 \sin (\theta ) \partial_r f
   \partial_r l r^4}{l}-4 r_H \kappa  \sin (\theta ) \partial_r V^2 r^3-8 \sin (\theta ) \partial_r f r^3+\frac{4 r_H \sin (\theta )
   \partial_r f \partial_r l r^3}{l}+4 \kappa  \omega_m \partial_r H_3 \partial_r V r^3\nonumber
   \\
   &-4 \kappa  \cos (2 \theta )
   \omega_m \partial_r H_3 \partial_r V r^3+4 \kappa  \csc (\theta ) g (-\omega  H_3 \sin (\theta )-m V)^2 r^2+4 \kappa  \sin
   (\theta ) \partial_\theta V^2 r^2+3 \kappa  \omega_m^2 \sin (\theta ) \partial_r H_3^2 r^2\nonumber
   \\
   &-\kappa  \omega_m^2 \sin (3 \theta ) \partial_r H_3^2 r^2-4
   \cos (\theta ) \partial_\theta f r^2-\frac{2 \sin (\theta ) \partial_\theta f \partial_\theta l r^2}{l}+12 r_H \sin (\theta )
   \partial_r f r^2-\frac{2 r_H^2 \sin (\theta ) \partial_r f \partial_r l r^2}{l}\nonumber
   \\
   &-4 r_H \kappa  \omega_m
   \partial_r H_3 \partial_r V r^2+4 r_H \kappa  \cos (2 \theta ) \omega_m \partial_r H_3 \partial_r V r^2-3 r_H \kappa  \omega_m^2 \sin (\theta ) \partial_r H_3^2 r+r_H \kappa  \omega_m^2 \sin (3 \theta ) \partial_r H_3^2 r\nonumber
   \\
   &+4 (r-r_H) \kappa  H_1^2 (-r \omega +m
   \omega_m)^2 \sin (\theta ) r+4 r_H \cos (\theta ) \partial_\theta f r+\frac{2 r_H \sin (\theta ) \partial_\theta f \partial_\theta l
   r}{l}+4 \kappa  H_3 \omega_m \sin (2 \theta ) \partial_\theta V r\nonumber
   \\
   &+4 \kappa  \omega_m \partial_\theta H_3 \partial_\theta V r-4
   \kappa  \cos (2 \theta ) \omega_m \partial_\theta H_3 \partial_\theta V r-4 r_H^2 \sin (\theta ) \partial_r f r-8 (r-r_H) \kappa  H_1
   (-r \omega +m \omega_m) \sin (\theta )\nonumber
   \\
   &\left(\omega_m \sin (\theta ) \partial_r H_3+r \partial_r V\right) r+3 \kappa  \omega_m^2 \sin (\theta )
   \partial_\theta H_3^2-\kappa  \omega_m^2 \sin (3 \theta ) \partial_\theta H_3^2+\kappa  H_3^2 \omega_m^2 \sin (\theta )+4 \kappa 
   H_2^2 (-r \omega +m \omega_m)^2 \sin (\theta )\nonumber
   \\
   &+\kappa  H_3^2 \omega_m^2 \sin (3 \theta )+2 \kappa  \cos (\theta ) H_3
   \omega_m^2 \partial_\theta H_3-2 \kappa  \cos (3 \theta ) H_3 \omega_m^2 \partial_\theta H_3-8 \kappa  H_2 (-r \omega +m
   \omega_m) \sin (\theta ) \left(\cos (\theta ) H_3 \omega_m\right.\nonumber
   \\
   &\left.\left.\left.+\sin (\theta ) \partial_\theta H_3 \omega_m+r \partial_\theta V\right)\right)\right)(r
   (r-r_H))^{-1}+\left(8 \left((r-r_H) \left(\partial_\theta f^2+r (r-r_H) \partial_r f^2\right)+l \left(2 \mu^2 \kappa  g
   V^2 r^3\right.\right.\right.\nonumber
   \\
   &-2 \omega_m \sin (\theta ) \left((r-r_H) \sin (\theta ) \partial_r \omega_m-2 \mu^2 r \kappa  g H_3 V(r,\theta
   )\right) r+\sin ^2(\theta ) \left(\partial_\theta \omega_m^2+r (r-r_H) \partial_r \omega_m^2\right) r\nonumber
   \\
   &\left.\left.\left.\left.+\left(2 \mu^2 r \kappa  g H_3^2+r-r_H\right) \omega_m^2 \sin ^2(\theta )\right)\right)\right)((r-r_H) f)^{-1}\right)
   %\\
    ,
\end{align}
\begin{align}
    \partial_\theta^2&g+r(r-r_H)\partial_r^2g=\frac{1}{2} \left(-\left(\kappa  f \left(6 r (r-r_H) \sin ^2(\theta ) \left(\partial_\theta H_1-\partial_r H_2\right)^2-g \left(2 m^2 r
   (r-r_H) H_1^2-4 m r (r-r_H) \sin (\theta ) \partial_r H_3 H_1 \right.\right.\right.\right.\nonumber
   \\
   &+2 m^2 H_2^2+\cos (2 \theta ) H_3^2+H_3^2-\cos (2 \theta ) \partial_\theta H_3^2+\partial_\theta H_3^2+r^2 \partial_r H_3^2-r r_H \partial_r H_3^2-r^2 \cos (2 \theta )
   \partial_r H_3^2+r r_H \cos (2 \theta ) \partial_r H_3^2\nonumber
   \\
   &\left.\left.\left.+2 H_3 \sin (2 \theta ) \partial_\theta H_3-4 m H_2 \left(\cos (\theta
   ) H_3+\sin (\theta ) \partial_\theta H_3\right)\right)\right) \csc ^2(\theta )\right)(r^2 l)^{-1}+6 \mu^2 \kappa  g^2 H_3^2+(r_H-2 r) \partial_r g\nonumber
   \\
   &+\frac{2 \left(\partial_\theta g^2+r (r-r_H) \partial_r g^2\right)}{g}\nonumber
   \\
   &+\frac{g
   \left(2 \mu^2 r (r_H-r) \kappa  H_1^2 l^2-2 \mu^2 \kappa  H_2^2 l^2+4 \cot (\theta ) \partial_\theta l
   l+4 r \partial_r l l-2 r_H \partial_r l l+\partial_\theta l^2+r^2 \partial_r l^2-r
   r_H \partial_r l^2\right)}{l^2}\nonumber
   \\
   &-\left(g \left((r-r_H) \left(\partial_\theta f^2+r (r-r_H) f^2\right)+l \left(2 \mu^2 \kappa  g V^2 r^3+2 \omega_m \sin (\theta ) \left(2 r \kappa  g H_3
   V \mu^2+3 (r-r_H) \sin (\theta ) \partial_r \omega_m\right) r\right.\right.\right.\nonumber
   \\
   &\left.\left.\left.-3 \sin ^2(\theta ) \left(\partial_\theta \omega_m^2+r (r-r_H) \partial_r \omega_m^2\right)
   r+\left(2 \mu^2 r \kappa  g H_3^2-3 r+3 r_H\right) \omega_m^2 \sin ^2(\theta )\right)\right)\right)((r-r_H) f^2)^{-1}\nonumber
   \\
   &+\left(2
   g \left(\kappa  \partial_r V^2 r^4-r_H \kappa  \partial_r V^2 r^3+2 \kappa  \omega_m \sin (\theta ) \partial_r H_3 \partial_r V
   r^3-3 \kappa  \omega ^2 g H_3^2 r^2-3 m^2 \kappa  \csc ^2(\theta ) g V^2 r^2+\kappa  \partial_\theta V^2 r^2\right.\right.\nonumber
   \\
   &+\kappa  \omega_m^2 \sin ^2(\theta ) \partial_r H_3^2 r^2-6 m \kappa  \omega  \csc (\theta ) g H_3 V r^2-r_H \partial_r f r^2-2 r_H
   \kappa  \omega_m \sin (\theta ) \partial_r H_3 \partial_r V r^2+(r-r_H) \kappa  H_1^2 (-r \omega +m \omega_m)^2 r\nonumber
   \\
   &-r_H \kappa 
   \omega_m^2 \sin ^2(\theta ) \partial_r H_3^2 r+2 \kappa  \cos (\theta ) H_3 \omega_m \partial_\theta V r+2 \kappa  \omega_m \sin
   (\theta ) \partial_\theta H_3 \partial_\theta V r+r_H^2 \partial_r f r-2 (r-r_H) \kappa  H_1 (-r \omega +m \omega_m)\nonumber
   \\
   &
   \left(\omega_m \sin (\theta ) \partial_r H_3+r \partial_r V\right) r+\kappa  \cos ^2(\theta ) H_3^2 \omega_m^2+\kappa  H_2^2
   (-r \omega +m \omega_m)^2+\kappa  \omega_m^2 \sin ^2(\theta ) \partial_\theta H_3^2+\kappa  H_3 \omega_m^2 \sin (2 \theta )
   \partial_\theta H_3\nonumber
   \\
   &\left.\left.\left.-2 \kappa  H_2 (-r \omega +m \omega_m) \left(\cos (\theta ) H_3 \omega_m+\sin (\theta ) \partial_\theta H_3
   \omega_m+r \partial_\theta V\right)\right)\right)(r (r-r_H) f)^{-1}\right)
   %\\
    ,
\end{align}
\begin{align}
   \partial_\theta^2&l+r(r-r_H)\partial_r^2l= \frac{2 \kappa  (r-r_H) f \left(\partial_\theta H_1-\partial_r H_2\right)^2}{r g}+\frac{2 \kappa  \mu^2 r g
   l^2 (\sin (\theta ) H_3 \omega_m+r V)^2}{(r-r_H) f^2}+\frac{2 \kappa  r g l
   (-\omega  H_3-m \csc (\theta ) V)^2}{(r-r_H) f}\nonumber
   \\
   &+\frac{-4 \kappa  \mu^2 g H_3^2 l^2+r^2
   \partial_r l^2+\partial_\theta l^2-r r_H \partial_r l^2+l \left(3 (r_H-2 r) \partial_r l-4 \cot (\theta )
   \partial_\theta l\right)}{2 l}
   %\\
    ,
\end{align}
\begin{align}
    \partial_\theta^2&\omega_m+r(r-r_H)\partial_r^2\omega_m=\frac{2 \left(\partial_\theta f \partial_\theta \omega_m+(r-r_H) \partial_r f \left(r \partial_r \omega_m-\omega_m\right)\right)}{f} +\left(\kappa  \csc ^2(\theta ) f \left(-2 r (r-r_H) H_1 \right.\right. \nonumber\\
    &\left(\sin (\theta ) \partial_r H_3 (2 m \omega_m-r \omega )+m r
   \partial_r V\right)+2 m r (r-r_H) H_1^2 (m \omega_m-r \omega )-2 H_2 \left(\sin (\theta ) \partial_\theta H_3 (2 m \omega_m-r \omega )\right.\nonumber
   \\
   &\left.+\cos (\theta ) H_3 (2 m \omega_m-r \omega )+m r \partial_\theta V\right)+2 m H_2^2 (m \omega_m-r \omega )+r^2
   \partial_r H_3^2 \omega_m-r^2 \cos (2 \theta ) \partial_r H_3^2 \omega_m+\partial_\theta H_3^2 \omega_m\nonumber
   \\
   &+2 \sin (2 \theta )
   H_3 \partial_\theta H_3 \omega_m-\cos (2 \theta ) \partial_\theta H_3^2 \omega_m-r r_H \partial_r H_3^2 \omega_m+r r_H \cos (2 \theta ) \partial_r H_3^2 \omega_m+2 r^3 \sin (\theta ) \partial_r H_3 \partial_r V\nonumber
   \\
   &\left.\left.-2 r^2 r_H \sin (\theta )
   \partial_r H_3 \partial_r V+2 r \sin (\theta ) \partial_\theta H_3 \partial_\theta V+H_3^2 \omega_m+\cos (2 \theta ) H_3^2
   \omega_m+2 r \cos (\theta ) H_3 \partial_\theta V\right)\right)(r^2 l)^{-1}\nonumber
   \\
   &+\frac{2 \omega_m \left(\kappa  \mu^2 r g
   H_3^2+r-r_H\right)}{r}+2 \kappa  \mu^2 r \csc (\theta ) g H_3 V-\frac{3 \left(\partial_\theta l
   \partial_\theta \omega_m+(r-r_H) \partial_r l \left(r \partial_r \omega_m-\omega_m\right)\right)}{2 l}\nonumber
   \\
   &-3 \cot (\theta )
   \partial_\theta \omega_m+2 (r_H-r) \partial_r \omega_m
    .
\end{align}
\end{widetext}
\subsection{Boundary conditions}

In order to solve the set of elliptic equations given above, we must impose boundary conditions,
which we list here for completeness. 
The $\theta$ boundary conditions follow from regularity at the poles and
from the assumed symmetry across the $\theta = \pi/2$ plane (equatorial symmetry),
which is used to reduce the domain to $0 < \theta < \pi/2$. 
At $\theta = 0$ the boundary conditions are
\begin{align*}
     \partial_{\theta}f=\partial_{\theta}l=\partial_{\theta}\omega_m=V&=H_1=\partial_{\theta}H_2=\partial_{\theta}H_3 =\phi = 0,\\
    g &= 1
     ,
\end{align*}
and at the $\theta = \pi/2$ boundary we have
\begin{align*}
     \partial_{\theta}f=\partial_{\theta}g=\partial_{\theta}l=\partial_{\theta}\omega_m=\partial_{\theta}V=\partial_{\theta}H_1\\=H_2=\partial_{\theta}H_3 =\partial_{\theta}\phi = 0
     .
\end{align*}

In the radial direction, the boundary at the black hole horizon can be derived by  
using Frobenius's method applied to the equations listed in Sec.
\ref{sec:listing_diff_equations} (along with the synchronicity condition), while those at spatial infinity ($x=1$) follow from the assumption
of asymptotic flatness.
For the $r = r_H$ (the black hole horizon) boundary, we find
\begin{align}
      \partial_{r}f= \partial_{r}g=\partial_{r}l=\partial_{r}&H_i= \partial_r \phi=0 , \\
      \omega_m = &m\Omega_H, \\   
      \left(V+\frac{\omega}{m}H_3 \sin\theta\right)& = 0
      ,
\end{align}
while at the $r = \infty$ boundary we find 
\begin{align*}
  f = g = l &= 1 ,\     & \omega_m = V &= H_i =\phi =0
  .
\end{align*}

In the scalar case, the above boundary 
conditions are equivalent to those in Ref. \cite{Herdeiro:2015gia}. 
In the vector case, the boundary conditions are equivalent to those
in Ref. \cite{Herdeiro:2016tmi}, with the exception of 
the condition on the $H_1$ matter field at the black hole's horizon: 
$\partial_rH_1(r_H) = 0$ [cf. Ref.~\cite{Herdeiro:2016tmi} which lists $H_1(r_H) = 0$]. 

\subsection{Seeding solutions}

We discretize the elliptic equations given above with fifth-order accurate
finite difference stencils, and solve them using Newton-Raphson relaxation. 
The relaxtion method requires a sufficiently good initial guess for the field
values in order to converge. 
In the scalar case, we use a numerical solution from Ref.~\cite{Herdeiro:2015gia} 
as an initial guess when solving for one point in the parameter space (and find a consistent solution), and then find subsequent 
solutions by slowly varying the parameters of the system while using the previous solution as
an initial guess.

In the vector case, we instead
\footnote{ 
Relativistic solutions for black hole-vector boson clouds
are given in Ref.~\cite{Santos:2020pmh}, though we
were unable to use them to relax 
to a solution to our equations, which could be related
to the difference in boundary conditions discussed above.
However, our solutions do seem equivalent in that the extracted black hole and cloud masses are equivalent for the same choice of frequency and black hole radius.
}
seed vector solutions using the analytic solution in the nonrelativistic limit
\cite{Baryakhtar:2017ngi} superimposed with a Kerr solution for the metric (see
Ref. \cite{Herdeiro:2015gia} Appendix A for a description of the Kerr metric in
these coordinates). In the nonrelativistic limit, the vector matter fields have
the solution:
\begin{align}
    V &= -\frac{\exp(-r\alpha)\alpha^{5/2}\sin(\theta)}{2\sqrt{\pi}\omega}
    &H_1 &= \frac{\exp(-r\alpha)\alpha^{3/2}\sin(\theta)}{2\sqrt{\pi}}\nonumber\\
    H_2 &= \frac{\exp(-r\alpha) r \alpha^{3/2}\cos(\theta)}{2\sqrt{\pi}}
    &H_3 &= \frac{\exp(-r\alpha) r \alpha^{3/2}}{2\sqrt{\pi}}
    .
\end{align}

We note that there are multiple solutions that satisfy the symmetry conditions
that we impose here. These solutions can be labeled by mode number $n$, where
the different behavior of the $n=0$ and $n=1$ solutions is discussed in
Ref.~\cite{Santos:2020pmh}. Here we are interested in the fundamental $n=0$ mode
(which grows the fastest through superradiance when $m=1$), which can be identified by the absence of nodes in
the $V$ component of the matter solution.

\section{Quantifying errors in frequency shift}
\label{app:errors}

In this section, we provide details on how the errors in the cloud mass-dependent shift in the oscillation
frequency are estimated.
There are multiple sources of error in the frequency shift calculation. The main sources are the following:
\begin{itemize}
    \item The complex field approximation
    \item Numerical error in individual solutions
    \item Interpolation error 
\end{itemize}

\begin{figure}[h]
\centering
\includegraphics[width= 8cm]{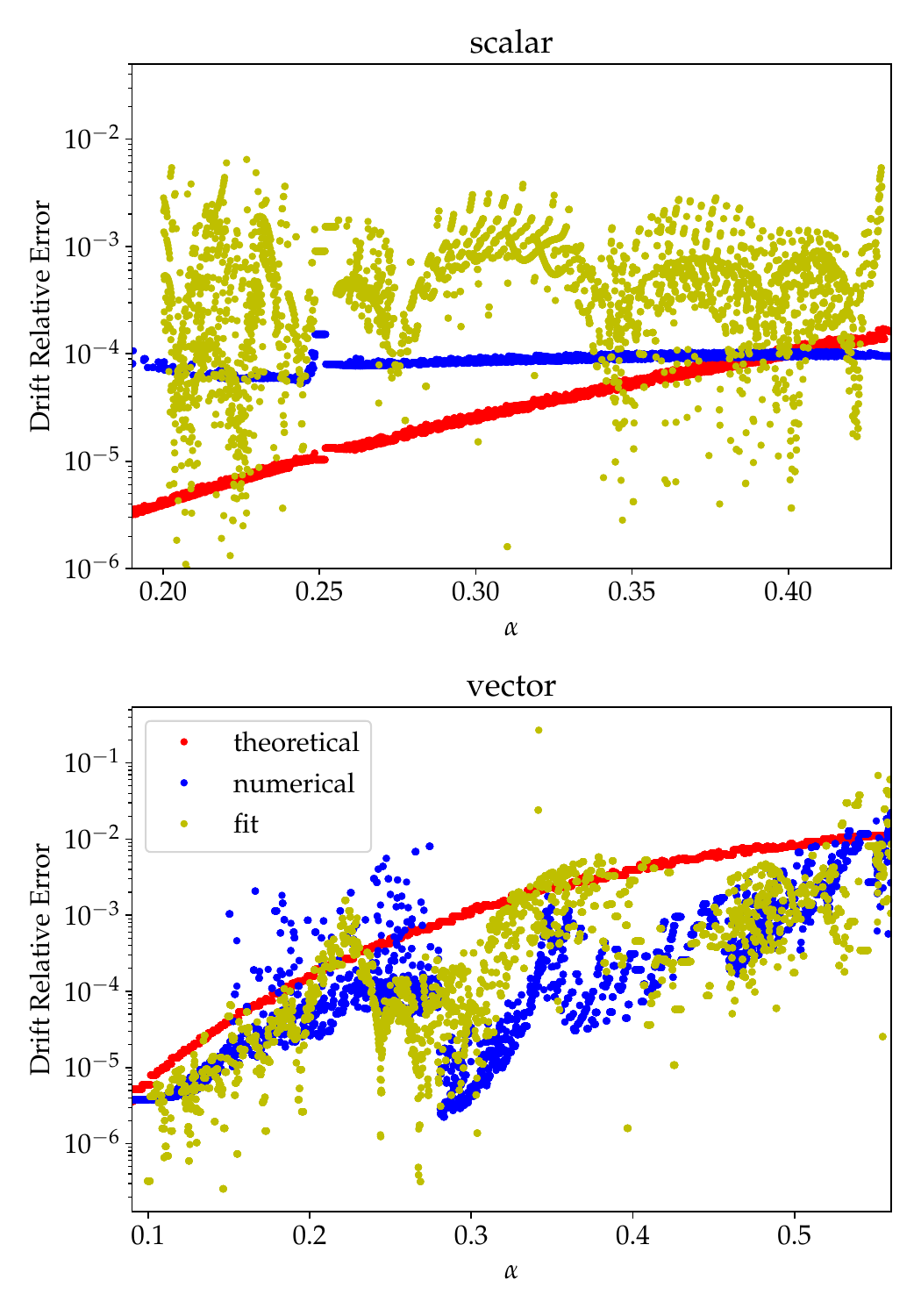}
\caption{Estimated errors in the extracted frequency shift relative to the
nonrelativistic limit of the frequency shift. We compare the final numerical
error in each solution to the estimated theoretical error due to our
oscillation averaged approximation, as well as the interpolation error.
Here, the numerical and interpolation errors are plotted for solutions with
$M_{\text{cloud}}/M_{\text{BH}}>1\%$ to avoid issues with relative error becoming larger as the frequency shift goes to zero. 
}
\label{fig:num_theo_errors}
\end{figure}

We take the overall relative error to be 
\begin{align}\label{eq:new_error}
    \frac{\delta\omega}{\Delta \omega} = \left|E_{\text{theoretical}}\right|+\left|E_{\text{numerical}}\right|+\left|E_{\text{fit}}\right|
    ,
\end{align}
where $E_{\text{theoretical}}$ is the relative error due to the complex field
or oscillation-average approximation, $E_{\text{numerical}}$ is the error from
the extraction of physical quantities from each numerical solution, and
$E_{\text{fit}}$ is the interpolation error. The results of these estimates are shown in Fig. \ref{fig:num_theo_errors}. We outline the calculation of each
of these contributions to the overall error in the following.

\subsection{Complex field approximation}
\label{sec:complex_field_approx}

\subsubsection{Comparing Newtonian limit term}
\label{sec:complex_err_newtonian_term}
As discussed in Sec.~\ref{sec:osc_avg}, we make an axisymmetric or oscillation-averaged
approximation to reduce the dimensionality of the problem. The theoretical error
arising from making such an approximation can be obtained by comparing the stress-energy
tensors of the full complex-field $\Phi$ solution to that of a real-field solution $\psi$.
To that end, we obtain a real-field solution by taking the real part of the complex field configuration
(adjusted by a factor of $\sqrt{2}$ to obtain a field with the same total energy), 
such that $\psi=\sqrt{2}\text{Re}(\Phi)$. In the vector case, we proceed in the same way, with $\mathcal{A}_a=\sqrt{2}\text{Re}(A_a)$.

Since the self-gravity correction scales with the binding energy of the cloud
\cite{Isi:2018pzk,Baryakhtar:2017ngi} (in the Newtonian limit), we will use that quantity to estimate
the theoretical error.
We introduce the quantity $W^{\ell\leq l'}$, defined as the $l'$ multipole approximation of the
nonrelativistic binding energy,
\begin{align}
    W^{\ell\leq l'} &= -\int d^3 \mathbf{r}\int_{|\mathbf{r'}|<|\mathbf{r}|}d^3\mathbf{r'} \rho(\mathbf{r})\rho(\mathbf{r'})\times\\ &\sum_{l=0}^{l'} \sum_{m=-l}^{l}\left( \frac{|\textbf{r}'|^{l}}{|\textbf{r}|^{l+1}}\right)\frac{4\pi}{2l+1}Y_{lm}(\Omega)Y^*_{lm}(\Omega')
    ,
\end{align}
which is obtained by applying a truncated version of the sum given in Eq.~\eqref{multipole} to the expression for $W$ given in Eq.~\eqref{eq:bindingE}.
In computing the above expressions, we use the energy density measured 
with respect to an observer whose four velocity
$n_a$ is normal to slices of constant time
\begin{align}
    \rho&= n_a n_b T^{ab}
    .
\end{align}
For the volume element, we take $d^3 \mathbf{r} = \sqrt{\gamma} d^3x$, 
where $\gamma$ is the determinant from the induced spatial metric in our coordinates [see Eq. \eqref{metric_ansatz}].
We then estimate the theoretical error to be
\begin{align}
    E_{\text{theoretical} }&\approx \frac{\left|W_{\text{Re}}^{\ell\leq 2} - W_{\text{C}}^{\ell\leq 2}\right|}{W_{\text{C}}^{\ell\leq 0}}
    ,
\label{eqn:eth}
\end{align}
where ``Re" and ``C" indicate the quantity computed with the 
stress-energy tensors corresponding to the real and complex solutions ($\psi$ and $\Phi$ in the scalar case, or $\mathcal{A}_a$ and $A_a$ in the vector case), respectively. 
In either case, the metric is taken to be identical and equal to that of the full
stationary and axisymmetric spacetime. 
This calculation estimates the error in a nonrelativistic approximation of the self-gravity of the cloud, the binding energy, calculated up to the $l=2$ multipole. We calculate the next nontrivial multipole in the scalar case ($l=3$) and find it to be subdominant to the $l=2$, $m=2$ contribution at the ratio of $10^{-5}$ at $\alpha=0.5$, indicating even higher multipoles are negligible. 

Using the estimate~\eqref{eqn:eth}, we obtain subpercent
level errors across the parameter space in the vector case,
and errors of at most $O(10^{-4})$ in the scalar
case, as can be seen in Fig. \ref{fig:num_theo_errors}. This error estimate is a nonrelativistic approximation in both spin
cases, and should only be fully accurate in that limit. Because it is a leading order
calculation, it will not capture the more quickly growing contributions at higher
orders in $\alpha$, and thus may underestimate the error due to this
approximation at high $\alpha$.
However, here we are mainly using it to estimate the relative contribution of nonaxisymmetric
components of the cloud's energy to the self-gravity. 
We further note that Ref.~\cite{Cannizzaro:2023jle} estimates the frequency shift for
the scalar case using relativistic black hole perturbation theory, but with a semi-Newtonian approximation
for the gravitational potential
and finds results noticeably closer to the relativistic ones presented here than to the leading order nonrelativistic result, suggesting that the frequency shift is well described by taking some leading order nonrelativistic terms. This further motivates us to use our nonrelativistic estimate for the theoretical error in taking an oscillation-averaged approximation, even at higher $\alpha$. 
As can be seen in Fig.~\ref{fig:num_theo_errors}, the theoretical error is comparable or subdominant to other sources of error in most of the parameter space, 
and so the exact value of the theoretical error here (up to a factor of a few) does not dramatically impact the total error estimate.

\subsubsection{Comparing to numerical relativity}
\label{sec:nr_comp}

\begin{figure}[h]
\centering
\includegraphics[width=0.5\textwidth]{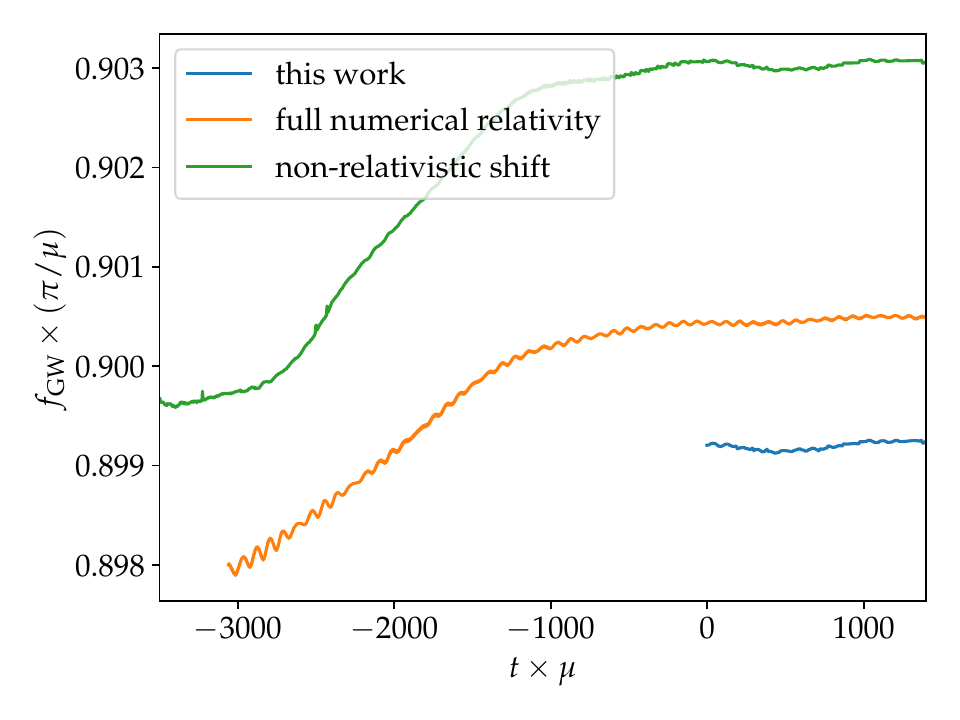}
    \caption{The frequency of gravitational waves  from an oscillating vector
    boson cloud around a black hole is compared in three cases. The numerical
    relativity result (averaged over a number of oscillations) is from a simulation including the growth period of the
    cloud from Ref. \cite{East:2017ovw}, and the other two results are using
    the test field frequency calculation from Ref. \cite{Siemonsen:2022yyf},
    and the leading order nonrelativistic frequency shift or the relativistic
    frequency shift as is calculated here. Here, the superradiant instability
    phase ends at $t=0$, after which this frequency shift calculation is
    valid.}
\label{fig:NR_comp}
\end{figure}

In addition to calculating a nonrelativistic estimate of the theoretical error
as is described in the last section, we can also compare to the numerical
relativity simulation considered in Ref. \cite{East:2018glu}.  That study
evolved the full Einstein-Proca equations, without symmetry assumptions,
through the saturation of the superradiant instability and into the phase where
the cloud dissipates through gravitational radiation for one case where
initially $\chi_{\rm BH}=0.99$ and $\alpha=0.4$, leading to a vector boson cloud
that peaks at $M_{\text{cloud}}\approx 0.06M_{\rm BH}$. We compare the gravitational
wave frequency found in that study to the test-field frequency with the nonrelativistic frequency shift and fully relativistic
frequency shift estimate (using the estimates of the black hole and cloud properties
found in the simulation) in Fig.~\ref{fig:NR_comp}.  We find an improved
agreement with numerical relativity, compared to the test nonrelativistic shift estimate, when
the relativistic frequency shift calculation is included in the waveform.
Considering the expected numerical error in the simulations, as well as the
fact that there is a small spurious excitation of the first overtone mode as
discussed in Ref.~\cite{East:2018glu}, which will bias the frequency upward,
the difference, at the level of $0.1\%$, is within the expected bounds.

\subsection{Numerical error in solutions}
\label{sec:numerical_error}
We quantify the numerical error $E_{\text{numerical}}$ using a combination of
errors in the extracted physical quantities. For the frequency fit, 
the frequency is a fixed input parameter, and
the only
extracted quantities are the black hole mass and the cloud mass (described in
Appendix \ref{sec:physical_quantities}). We approximate the relative numerical error in a particular solution
by the sum of the relative errors in both the cloud and black hole mass. 

We establish the numerical errors in each mass calculation by performing
convergence studies. The elliptic solver uses sixth order finite differencing
and so the solution in the bulk converges at sixth order. To calculate the mass
of the black hole we sometimes use an extrapolation to the horizon point (as
described in Appendix \ref{sec:physical_quantities}), and so have slower
convergence.

In the scalar case, we perform a convergence test using three resolutions
differing by a factor of $2/3$. We find second order convergence in the horizon
black hole mass and sixth order convergence in the cloud mass, and we perform
Richardson extrapolation to find an overall error. In this case, the black hole
mass dominates the numerical error at small $\alpha$. The reason the black hole
mass converges so slowly is because its calculation relies on an extrapolation
to the boundary, where the extrapolation error is dominating over the error in
each point in the bulk. For the final fit, we use a resolution of $n_x=391 $ radial points
and $n_{\theta}=46$ angular points. The result of this analysis is shown in comparison with the
other sources of error in the top panel in Fig. \ref{fig:num_theo_errors}. 

In the vector case 
we perform a convergence test using three resolutions differing by a factor of
$3/4$ (where our numerical resolution of $n_x=261 $ and $n_{\theta}=35$ is the
highest resolution). We find sixth order convergence in the ADM mass and cloud
mass, with second order convergence in the horizon black hole mass. We perform
Richardson extrapolation to find an overall error. In this case the horizon
black hole mass has good convergence at low $\alpha$, but stops being well
resolved at high $\alpha$ ($\alpha> 0.4$), whereas the ADM mass converges well
at high $\alpha$, but stops converging at these resolutions when the cloud
becomes bigger, and closer to the compactified boundary, at $\alpha
<0.2$\footnote{We check the convergence using a similar study but where the
minimum resolution is $n_x=347 $ and $n_{\theta}=47$, and find that the ADM
mass converges as expected for $\alpha>0.13$.}. For $\alpha>0.22$,
we use a resolution of $n_x=261 $ and $n_{\theta}=35$, and the ADM and cloud
mass calculations. We use the horizon black hole mass instead of the ADM mass
as well as higher resolution points for $\alpha < 0.22$, with varying
resolution so as to make the estimated numerical errors comparable to the
theoretical error in that regime. The final error estimates are shown in Fig. \ref{fig:num_theo_errors}.
We perform a fit for the frequency shift for three different
resolutions of points, and find convergence in the overall fit parameters for the frequency
shift as well as in each solution.

There is also some contribution to the overall error from the interpolation
between solutions. We evaluate this error $E_{\text{fit}}$ by removing each
point in turn from the interpolation and calculating the difference between the
interpolation and that point. We show the results of this test for the scalar
and vector cases in Fig. \ref{fig:num_theo_errors}.

\subsection{Extrapolation to nonrelativistic solutions}
\label{sec:extrap_errors}

As described in Sec. \ref{sec:methods}, to extrapolate to small $\alpha$ we
fit the self gravity correction to the cloud frequency using a polynomial in
$M_{\text{cloud}}/M_{\text{BH}}$ and $\alpha$. We use points with $0.2<\alpha<0.3$ in the scalar case and $0.09<\alpha<0.14$ in the vector case. By construction, this polynomial reduces to the nonrelativistic result to leading order in the $\alpha\ll 1$ limit. It only remains to quantify the error in the polynomial fit. We do this by first calculating the maximum deviation of the polynomial fit from the full spline interpolation of the data points in the fit region, $\Delta_{\text{fit}}$. Then the maximum error in the coefficient for the leading order relativistic correction to the shift is $\delta c_{3,1} = \Delta_{\text{fit}} \alpha^{-3}$. So we can extrapolate the relative error in the frequency shift to lower values of $\alpha$ using 
\begin{align}
    \frac{\delta \omega}{\Delta \omega} &=  \frac{\delta c_{3,1}\alpha^3}{C_{\text{nr}}\alpha^2} = \frac{\Delta_{\text{fit}} \alpha}{C_{\text{nr}}}
\end{align}
for $\alpha$ below the regime where we have numerical results.
In the each spin case we find that the polynomial fit to the data is close to the numerical error for the nearby small $\alpha$ points.

\section{Calculating light ring and ISCO}
\label{sec:appendix_lr_isco}

We find the equatorial light ring by solving the radial geodesic equation
together with the normalization of the four velocity. We interpolate our
numerical solution between grid points before solving for the radius that gives
circular orbits\footnote{We use a cubic spline interpolation and an
optimization algorithm from \texttt{scipy}.}. We use the proper circumferential radius to
convert our measured light ring to standard coordinates
\begin{align}
    R &= \frac{1}{2\pi} \int d\varphi \sqrt{g_{\varphi\varphi}(r,\theta=\pi/2)} .
\end{align}

We calculate the equatorial ISCO by solving the radial geodesic equation,
varying the conserved quantities $E = -t_a u^a$ and $L = \phi_a u^a$ (where
$u^a$ is a timelike geodesic four velocity, $t^a$ is the stationary Killing
vector and $\phi^a$ is the axial Killing vector) and the estimated ISCO radius
to minimize simultaneously $\Dot{r}$, $\partial \Dot{r}/\partial r$,
and $\partial ^2\Dot{r}/\partial r^2$. The resulting solution gives the ISCO radius, as well as $E$
and $L$ which can be mapped to the four velocity $u^a$ of a particle traveling
along the ISCO. This four velocity can then be used to calculate the ISCO
orbital frequency as well as the frequency of light measured by that observer.
We use the proper time measured by an observer orbiting on the ISCO to define a coordinate
independent orbit frequency,
\begin{align}
    \Omega &= \frac{d\varphi}{d\tau} .
\end{align}
The frequency of light measured by that observer is $\omega = -u^ak_a$, where
$k^a$ is the four velocity of the light being measured. We choose two initial
directions for light $k^a$ being emitted from the ISCO, such that in both cases
the light is emitted equatorially and the initial radial component is zero. 

\begin{figure}
    \centering   
    \subfloat[\centering Scalar field]{{\includegraphics[width=8cm]{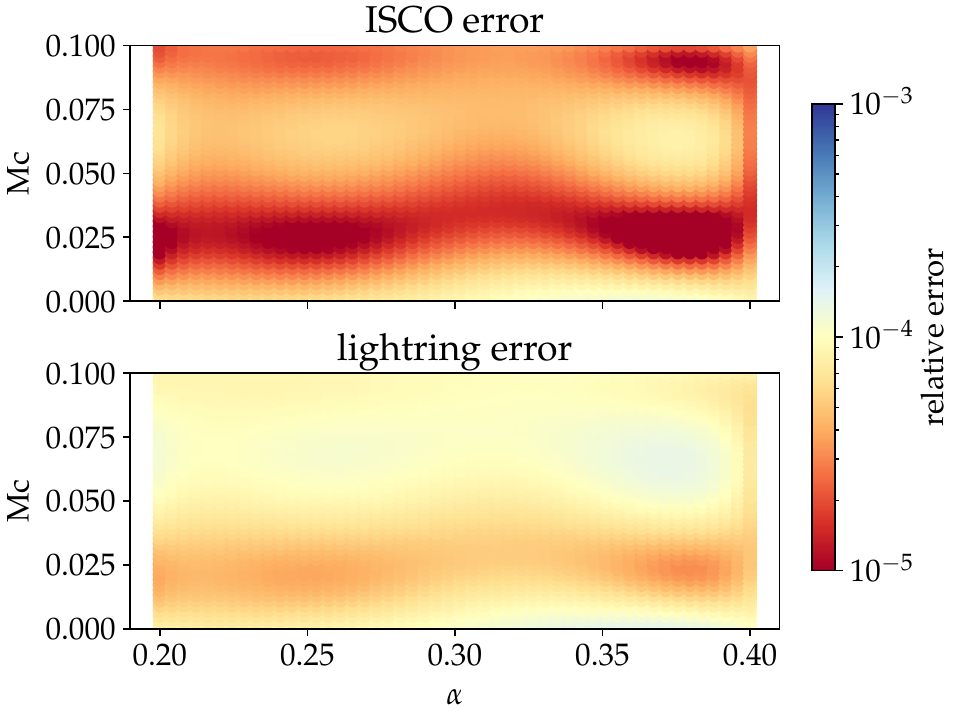} \label{fig:error_lrisco_a}}}%
    \qquad
    \subfloat[\centering Vector field]{{\includegraphics[width=8cm]{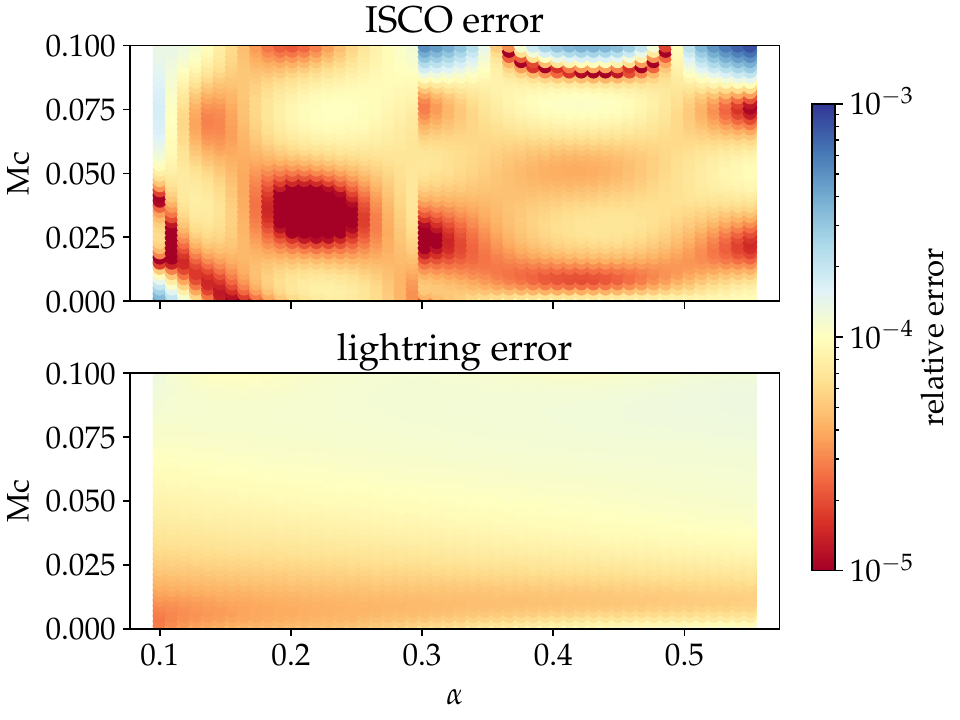} \label{fig:error_lrisco_b}}}%
    \caption{We perform a convergence test to determine the numerical error in the position of the light ring and ISCO. The relative error in the circumferential radius is shown here for both scalar and vector clouds. The feature at $\alpha = 0.3$ in the ISCO radius for the vector case is due to a change in the convergence order from 6th order at high $\alpha$ to second order for low $\alpha$.
    }%
    \label{fig:lr_isco_errs}
\end{figure}

Errors in the light ring and ISCO calculations are evaluated using a
convergence test and Richardson extrapolation, with results shown in Fig.
\ref{fig:lr_isco_errs}. As can be seen there, the relative errors are always
$<10^{-3}$ and are $\lesssim 10^{-4}$ except in some corner of the vector boson
parameter space.

\section{Consistency checks}
\label{sec:consistency_checks}

\subsection{Test field limit}

\begin{figure}[h]
\centering
\includegraphics[width=0.5\textwidth]{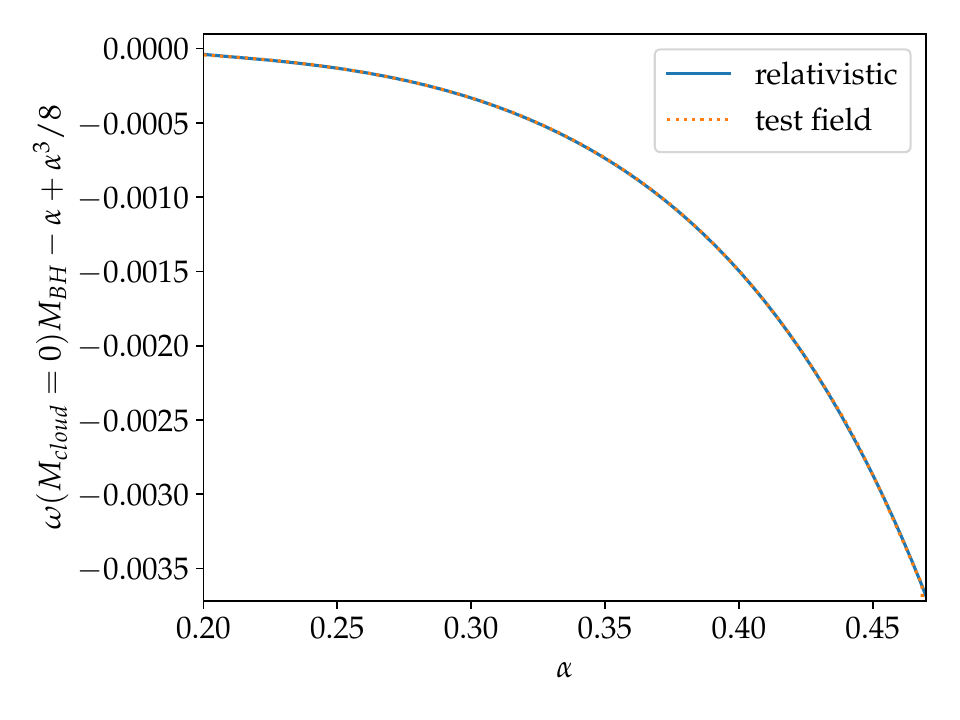}
\caption{We show the relativistic correction to the frequency in the test field
limit and confirm that our result matches that from the test field calculations
in Ref.~\cite{Siemonsen:2022yyf}. This example is from the scalar case, but we see a
similar agreement in the vector case.
}
\label{fig:test_field_lim}
\end{figure}

We confirm that in the test field limit our results match those in the
literature. We do this by comparing our fit to the constant in cloud mass part
of the frequency with the test field calculation in Ref.
\cite{Siemonsen:2022yyf}. The results of this comparison are shown in Fig.
\ref{fig:test_field_lim}. We find agreement within the expected error bars of
our calculation.

\subsection{Nonlinear dependence of frequency on $M_{\text{cloud}}$}
\label{sec:higher_order_mc_fit}

\begin{figure}
    \centering   
    \subfloat[\centering Scalar field]{{\includegraphics[width=8cm]{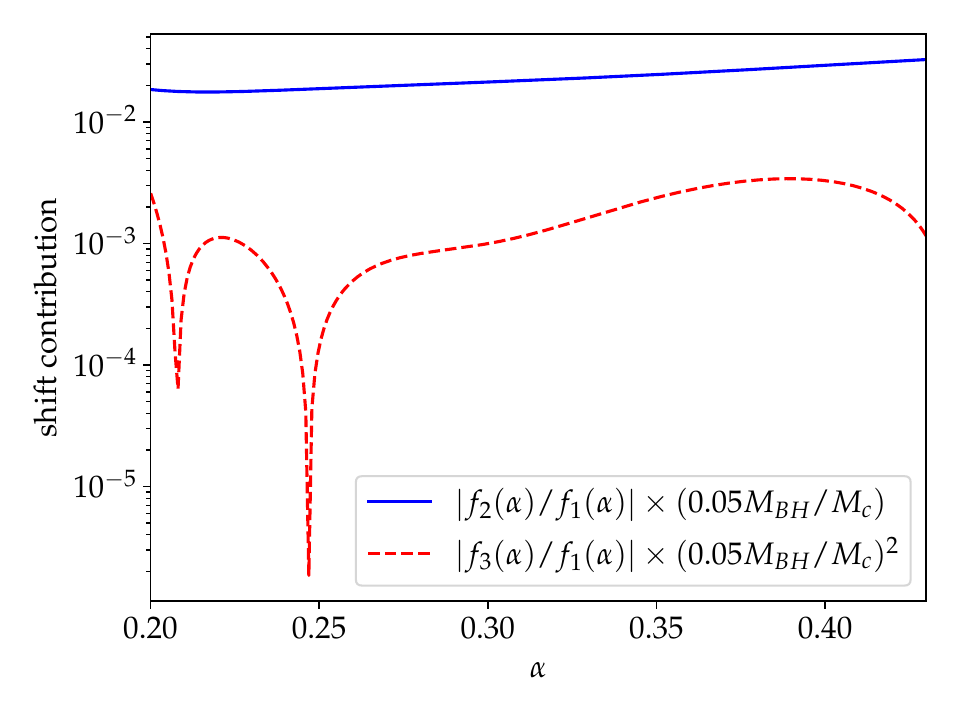} \label{fig:Mc3_a}}}%
    \qquad
    \subfloat[\centering Vector field]{{\includegraphics[width=8cm]{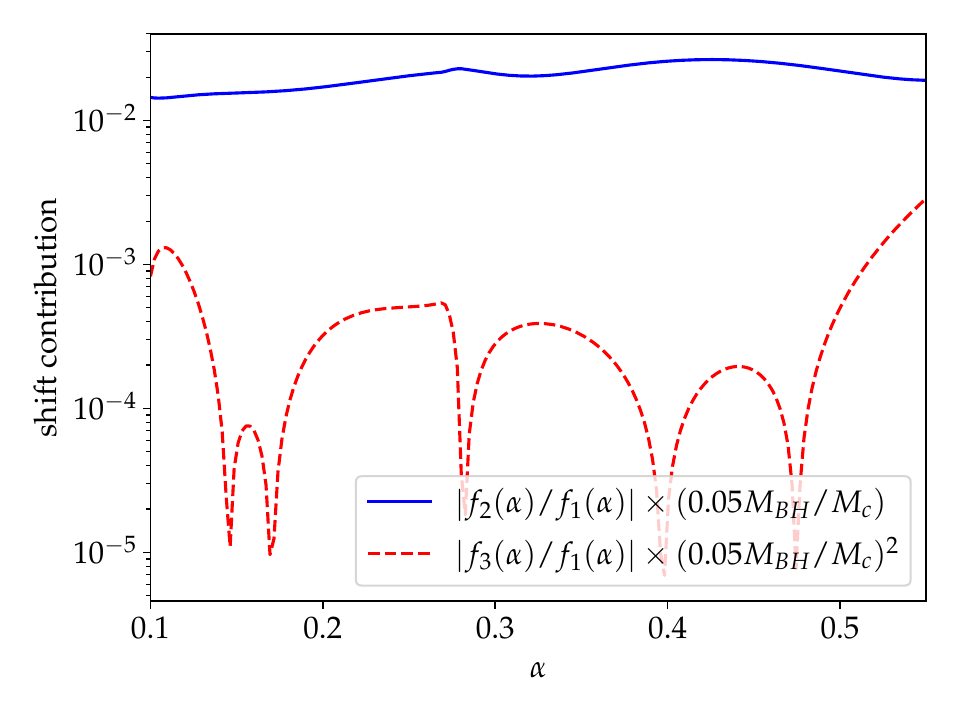} \label{fig:Mc3_b}}}%
    \caption{The contribution to the frequency shift of the second order in
cloud mass term, relative to linear term, when the cloud mass is $M_{\text{cloud}}/M_{\text{BH}}= 0.05$.
We also include a comparison for the value of a third order term when it is included in the
fit described in Eq.~\eqref{eq:full_numeric_fit}. When this third order term is
included, it is found to contribute at below the estimated error. 
The feature at $\alpha = 0.275$ in panel (b) is due to
an interpolation between fits to data at different resolutions.}
    \label{fig:Mc3_contribution}
\end{figure}

As described in Sec. \ref{sec:fitting_data}, when fitting for the cloud mass
dependent part of the frequency, we include a term linear in cloud mass as well
as a quadratic term in the fit to the frequency shift. As shown in Fig.~\ref{fig:Mc3_contribution},
the quadratic term corrects the linear term only at the level of a few percent, 
even when the cloud is significant. 
We find that including an additional cubic in cloud mass term does not significantly decrease the root mean squared
error in the fit, and, as is shown in Fig. \ref{fig:Mc3_contribution}, that
term is oscillating around zero and does not appear to be well resolved.

\bibliography{bib.bib}
\end{document}